\documentstyle[preprint,aps,epsfig]{revtex}
\begin{document}
\tighten  
\title{Nuclear effects in $g_{1A}(x,Q^2)$ at small $x$ in deep inelastic scattering on $^7$Li and $^3$He}

\author{V.~Guzey}
\address{Department of Physics, The Pennsylvania State University, University Park, PA 16802}
\author{M.~Strikman}
\address{Department of Physics, The Pennsylvania
State University, University Park, PA 16802, USA,\\
and Deutsches Elektronen-Synchrotron (DESY), Germany\thanks{On leave of
absence from PSU.}}

\maketitle

\begin{abstract}

We suggest to use polarized nuclear targets of $^7$Li and $^3$He to study  nuclear effects in the spin dependent structure functions $g_{1A}(x,Q^2)$. These effects are expected to be enhanced by a factor of two as compared to the unpolarized targets.
 We predict a significant $x$ dependence at $10^{-4} \div 10^{-3} \leq x \leq 0.2$ of $g_{1A}(x,Q^2)/g_{1N}(x,Q^2)$ due to nuclear shadowing and nuclear 
enhancement. The effect of nuclear shadowing at $x \approx 10^{-3}$ is of an order of $16\%$ for $g_{1A=7}^{n.s.\,3/2}(x,Q^2)/g_{1N}^{n.s.}(x,Q^2)$ and $10\%$  for $g_{1A=3}^{n.s}(x,Q^2)/g_{1N}^{n.s.}(x,Q^2)$. By imposing the requirement that the Bjorken sum rule is satisfied
 we model the effect of enhancement.
 We find the effect of  enhancement at $x \approx 0.125\ (0.15)$ to be of an order of $20\ (55)\%$ for $g_{1A=7}^{n.s.\,3/2}(x,Q^2)/g_{1N}^{n.s.}(x,Q^2)$  and $14\ (40)\%$ for $g_{1A=3}^{n.s}(x,Q^2)/g_{1N}^{n.s.}(x,Q^2)$, if enhancement occupies the region $0.05 \leq x \leq 0.2$ ($0.1 \leq x \leq 0.2$). We predict  a 2\% effect in the difference of the scattering cross sections  of deep inelastic scattering of an unpolarized projectile off $^7$Li with $M_{J}$=3/2 and $M_{J}$=1/2. We also show explicitly that the many-nucleon description of deep inelastic scattering off $^7$Li becomes invalid in the enhancement region $0.05 < x \leq 0.2$.

\end{abstract}

\section{Introduction}
\label{intro}

It is now well known
 that nuclear effects are important in deep inelastic scattering (DIS) 
of leptons on nuclear targets. For example, if  we consider the ratio of structure functions  $R_{A}(x,Q)=\frac{F_{2A}(x,Q)/A}{F_{2D}(x,Q)/2}$ as a function 
of $x$, we can distinguish various regions over the range of $x$ from 
$10^{-5}\div10^{-4}$ to $x=1$ and beyond, where different nuclear
 mechanisms govern  particular behavior of $R_{A}(x,Q)$. 
In this paper we will consider only the region $x \leq 0.2$.

For $x < 0.05$ $R_{A}(x,Q)$
 is smaller than 1 due to nuclear shadowing. It arises from
  coherent interactions of the incoming virtual photon with several
 nucleons of the target nucleus in the target rest frame. 
These interactions occur over the distance $l_{c}=t_{f} \sim 1/(2m_{N}x)$, see Eq.\ (\ref{cond}). When $l_{c} \sim r_{NN}=1.7$ fm, the average internucleon distance in nuclei, the coherence effects become suppressed. This sets the upper limit to the region of nuclear shadowing  $x \approx 0.05$.
 In the infinite momentum frame the leading twist shadowing can
 be viewed      as  a depletion of the nuclear   parton
 densities due to a   spatial overlap of partons belonging to different nucleons.
Nuclear shadowing phenomena 
for the structure function $F_{2A}(x,Q^2)$ and for the nuclear gluon
parton density  $G_A(x,Q^2)$
have been discussed in the literature for a long time. Current data on  shadowing for $F_{2A}(x,Q^2)$ are consistent with 
the Gribov theory \cite{gribov} which relates this phenomenon to diffraction 
on protons,
 for the recent discussion and references see \cite{FS981}.

For $x >$ 0.05 up to 0.2$\div$0.25 $R_{A}(x,Q^2)$ first grows
 and becomes larger than 1, then decreases and approaches 1 from above. The   dynamical mechanism of  enhancement has not been understood yet. 
It is present    for  the gluon and valence quark parton densities
\cite{FS88,FSL,Pirner}
 and it is absent for 
the sea quarks \cite{DY}.

Thus, nuclear shadowing and  enhancement are the two phenomena that
 modify the structure functions of nuclei at $10^{-5} \div 10^{-4}
 \leq x \leq 0.2 \div 0.25$ and produce their nontrivial $x$ dependence. 
In spite of the potential importance of the nuclear shadowing and 
enhancement effects 
until recently they were ignored in the studies of nuclear effects in 
 polarized  nuclear structure functions \cite{sf} 
and their applications
 to  the Bjorken sum rule for nuclei \cite{bsr}.

The study of nuclear shadowing provides  unique information
about the space-time picture of strong interactions
and, in particular, about the  interplay of soft and hard effects. The  uncertainties
due to the wave functions of nuclei are 
 small since the large
internucleon distances give the dominant contribution.

Another application of studies on nuclear shadowing effects
 at low $x$ is related to the 
neutron spin dependent structure function. 
As it  was pointed out in our original paper \cite{FGS}, when
 one extracts the neutron spin dependent structure function $g_{1n}(x,Q^2)$
 from the $^3$He data, shadowing and enhancement should be taken into consideration.
 If it is not done, this might lead to a misinterpretation of the data.

We propose to study experimentally effects of nuclear shadowing and 
enhancement  using  polarized 
nuclei of $^7$Li and $^3$He because  for these  nuclei  nuclear shadowing and 
enhancement are larger by a factor of {\bf two} as compared 
to the unpolarized targets \cite{FGS}
\footnote{A similar enhancement is present for deuteron targets,
though it is very difficult to observe it  experimentally
since $g_{1d}(x,Q^2) \ll g_{1p}(x,Q^2)$ for small $x$.
For an extensive discussion of  scattering off  polarized deuteron targets 
see \cite{Piller}.
}.

Basically, this is because in the case
of unpolarized scattering due to  $N$ identical exchanges
a factor $1/N!$ is present in the Glauber series, while in the case
of polarized scattering
the exchange generating $g_{1A}(x,Q^2)$ is not identical 
to
the
rest of the exchanges, leading to a factor $1/(N-1)!$.
Hence, for the case of light nuclei where triple rescatterings can be neglected
the shadowing effect is enhanced by a factor of two.
Also it is instructive to consider the limit of large atomic number $A$. In this case
scattering off a polarized valence nucleon can contribute to 
$g_{1A}(x,Q^2)$
only if the nucleon is near the edge since for the central impact parameters the interaction
would be black and, hence, would  not contribute to the spin asymmetry.
 As a result in this
 limit one would get for small enough $x$ 
that
 $g_{1A}(x,Q^2) \propto A^{-2/3} g_{1N}(x,Q^2)$.

Polarized $^3$He has been used extensively over years.
It seems feasible  to reach
large values of polarization for a $^7$Li target as well \cite{crabb}.

In the present paper  we calculated nuclear shadowing of the spin dependent
 structure functions $g_{1A}^{n.s.\,3/2}(x,Q^2)$ and $g_{1A}^{n.s.\,1/2}(x,Q^2)$ 
using an extension of the Gribov theory of nuclear shadowing in DIS \cite{gribov}.

 Obviously this changes the contribution
 to the Bjorken sum rule at small $x$. Hence this depletion should
 be compensated by some enhancement such that these two effects do
 not alter
the ratio of the integrated nuclear to nucleon non-singlet spin 
dependent structure functions, $R$, which is determined by the Bjorken sum rule. 
We used this as a guiding principle to model the effect of 
enhancement for the region $0.05\ (0.1) \leq x \leq 0.2$, 
where enhancement is the dominant nuclear effect.

\section{Nuclear effects at small $x$ in $g_{1A}(x,Q^2)$ for $^7$Li}
\label{lithium}

As we pointed out in the Introduction the fact that it is possible to create 
targets of polarized $^7$Li makes this nucleus a useful tool for studying spin 
dependent structure functions. According to the shell model nuclear 
polarization will be predominantly (87$\%$)  due to a single proton \cite{Landau}. 
This  will enhance shadowing and enhancement effects for the polarized target as 
compared to the unpolarized one by a factor of two. 
An advantage of using $^7$Li as compared to $^3$He, i.e. a nucleus with the 
''valence'' (in the sense of the shell model) proton rather than neutron, is
 that  the proton spin dependent structure function $g_{1p}(x,Q^2)$ is known
 with a better accuracy than the neutron spin dependent structure function 
 $g_{1n}(x,Q^2)$. This matters because  one can see from the structure of
 our answer that we compute the ratio of nuclear to nucleon spin dependent
 structure functions. Thus, in order to find the nuclear structure function
 alone, we need to know the nucleon structure function with as the best 
accuracy as possible.

It was realized long time ago that if the transition time of a photon
 with a four-momentum $Q$ into a quark-gluon configuration with mass $M_{h_{i}}$
is larger than the typical interaction time, which is of the size of the target,
 i.e.  when 
\begin{equation}
t_{f}=\frac{2\nu}{Q^2+M_{h_{i}}^2} \approx \frac{1}{2m_{N}x} \ge 2R_{A} \ ,
\label{cond}
\end{equation}
then the incoming virtual photon (could be a real one as well) reveals
 its hadronic properties. For example, DIS on nuclei displays shadowing,
 i.e. the effect, when the  nuclear structure function per nucleon  is 
smaller than the nucleon structure function.
Throughout this paper we will refer to these quark-gluon configurations 
as states 
$|h_{i}\rangle$.

If Eq.\ (\ref{cond})  is satisfied, then the total photoabsorption cross
 section of photons on a nuclear target with the atomic  number $A$ can
 be presented
\begin{equation}
\sigma_{\gamma^{\ast}A}(Q^2)=\sum_{h_{i}}|\langle \gamma^{\ast}|h_{i} \rangle|^2 \sigma_{h_{i}A} \ .
\label{cond2}
\end{equation} 

Note that since Eq.\ (\ref{cond2}) is an effective one, we can think 
of states $|h_{i} \rangle$ as of
 eigenstates of the scattering matrix, i.e. no non-diagonal 
transitions need to be introduced.

To elaborate more on on the cross sections $\sigma_{h_{i}N}$ and $\sigma_{h_{i}A}$ and their relevance to our problem it is instructive to go to the infinite momentum frame of the target. The spin dependent structure function $g_{1}(x,Q^2)$ can be written as a convolution with quark non-singlet $\Delta q_{ns}$, singlet  $\Delta q_{s}$ and gluon $\Delta g$ polarized parton densities \cite{ABFR}
\begin{equation}
g_{1}(x,Q^2)=\frac{\langle e^2 \rangle}{2}\Big(C_{ns}\otimes \Delta q_{ns}+C_{S}\otimes\Delta q_{s}+2n_{f}C_{g} \otimes \Delta G \Big) \ ,
\end{equation}
where $ \langle e^2 \rangle=n_{f}^{-1} \sum_{i=1}^{n_{f}}e_{i}^2$, $\otimes$ denotes convolution with respect to $x$. Therefore, nuclear shadowing of $g_{1A}(x,Q^2)$ at small $x$ is due to nuclear shadowing of the spin dependent structure functions. The recent analysis of the diffractive data of HERA \cite{FS981} showed that in unpolarized deep inelastic scattering on nuclei the nuclear shadowing for gluon parton densities should be larger by a factor of three than that for the quark parton densities. Namely, at $x \approx =10^{-3}$ and 
$Q$ is of an order of few GeV the effective cross section of interaction of $|h_{i} \rangle$ with the nucleon in the quark channel $\sigma_{eff}$=17 mbarn, and in the gluon channel  $\sigma_{eff}$=50 mbarn. The notion of the quark channel is referred to the $\gamma^{\ast}$ interaction with the quark field of the target and the gluon channel is referred to the interaction with the gluon field of the target.  
 
 Hence, we make {\it a natural hypothesis}
that the strengths of interactions in the sea quark channel of the unpolarized DIS and polarized channels are similar.
Since the shadowing in the quark channel is characterized by
a relatively small average interaction strengths it is a good approximation to
 replace the sum over hadronic components  of the photon  by a single
 term  with the typical mass  $M_{h}=Q$ and the cross
 section  $\sigma_{hN}=\sigma_{eff}$=17 mbarn.
This choice corresponds to the analysis of shadowing for nuclei
 with $A \ge 12$ in color screening models, where the fit to 
the experimental data 
at $x \sim 10^{-3} \div 10^{-2}$, $Q^2 \sim 2 \div 5$ GeV$^2$
required 
$\sigma_{eff}=\frac{\langle \sigma^2 \rangle}{\langle
 \sigma \rangle}$=17 mbarn (see \cite{FGS} and references therein). 
Here the averaging occurs with the measure of the probability of the corresponding
configurations $\left|h_{i} \right \rangle$
in the photon wave function.
The HERA diffractive $e\,p$  data
 lead to a similar value of $\sigma_{eff} \approx 14 \pm 3.5$ mbarn
 at somewhat smaller $x$ and larger $Q^2$.

A recent global NLO QCD analysis \cite{ABFR} of the world data from CERN, SLAC and DESY has showed that the present accuracy of the data is not sufficient to make any qualitative conclusions about the magnitude of the polarization of gluons $\Delta G(x,Q^2)$. Hence, we will disregard the contribution of $\Delta G_{A}(x,Q^2)$ to $g_{1A}(x,Q^2)$ in this paper.
At the same time 
 we will estimate the amount of nuclear shadowing in 
$\Delta G_{A}(x,Q^2)$ for $^3$He
which turns out to be large and may be possible to investigate at HERA in
 the polarized  eN mode.

In the discussed approximation 
the problem of deep inelastic scattering
 of a virtual photon is reduced 
to  scattering of an effective hadronic (quark-gluon) state. The latter can be 
treated using the Gribov-Glauber formalism.

Note that due to cross section fluctuations in the incoming photon the 
triple scattering term should contain $\langle \sigma^3 \rangle$, where one
 averages over fluctuations of the size of the projectile. Our assumption that
 the incoming photon interacts with the target through the effective  state
 $|h_{i} \rangle$ means that we have replaced $\langle \sigma^3 \rangle$ by 
$(\langle \sigma^2 \rangle)^2/ \langle \sigma \rangle=\sigma_{eff}^2 \langle \sigma \rangle$. 
Since the triple  term is numerically small  this substitution is
 of high accuracy.

Note that in difference from the hadron-nucleus scattering one 
has to take   
into account  a non-vanishing longitudinal 
momentum $q_{\parallel}$ transferred to the target in the transition $\gamma \rightarrow 
\left |\right. h\rangle$,  
$q_{\parallel}=\frac{Q^2+M_{h}^2}{2\nu} \approx 2m_{N}x$, see e.g. \cite{VDM}.

 In our calculation  we used the ground-state wave function of $^7$Li given 
by the nuclear shell model \cite{Landau}. 
This is certainly an oversimplified model for the $^7$Li wave function.
However, the shadowing effects are mostly determined by the long range
part of the
wave function and should not be sensitive
to the refinements of the wave function.

The impulse approximation is certainly more sensitive to the degree of the nucleon polarization in the nucleus than to the effects of the Fermi motion, which are small for $x \ge 0.5$.
So, in the kinematics that we discuss one can write
in the impulse approximation \cite{sf}
\begin{equation}
g_{1A}(x,Q^2)=P_p g_{1p}(x,Q^2)+ P_n g_{1n}(x,Q^2) \ ,
\label{impulse}
\end{equation}
where 
$P_{p}$ ($P_{n}$) is  polarization of the proton (the neutron) in $^7$Li.

We find that the modification of $g_{1A}(x,Q^2)$  due to
the nuclear shadowing cannot not be mimiced
by any reasonable variation of $P_{p}$ and $P_n$, see discussion in section\ \ref{manynucleon}.

In most of our  analysis of this section the $z$ component of 
the  total angular momentum of $^7$Li, $M_{J}$, is chosen to be 3/2 or -3/2.
 The axis of quantization is along the 
virtual photon
direction. Therefore, 
target polarization discussed in our paper is always longitudinal.
We will show that when the target is longitudinally 
polarized with $M_{J}$=1/2 the discussed effects are suppressed.
 For the  components of the wave function of $^7$Li with 
$M_{J}$=3/2, 1/2, -1/2, -3/2 see Appendix A.

 The  valence part of the total wave function of $^7$Li consists of two valence neutrons and one proton in the $1p_{3/2}$ shell. It is given by \cite{Landau}
\begin{eqnarray}
\Psi_{Li-7}^{3/2}&=&\frac{3}{\sqrt{15}}[\Psi_{p}^{3/2}\Psi_{n}^{3/2}\Psi_{n}^{-3/2}]-\frac{2}{\sqrt{15}}[\Psi_{p}^{3/2}\Psi_{n}^{1/2}\Psi_{n}^{-1/2}] \nonumber\\
&-&\frac{1}{\sqrt{15}}[\Psi_{p}^{1/2}\Psi_{n}^{3/2}\Psi_{n}^{-1/2}]+\frac{1}{\sqrt{15}}[\Psi_{p}^{-1/2}\Psi_{n}^{3/2}\Psi_{n}^{1/2}] \ .
\label{wf32text}
\end{eqnarray}

The superscripts refer to the $z$ component of the total angular momentum.
 $[\dots]$ stands for antisymmetrization. Four remaining nucleons occupy
 the $1s_{1/2}$ shell and their wave function is trivial. Being in the 
symmetric $S$-state they do not contribute to the spin asymmetry.  
 This nuclear wave function\ (\ref{wf32text}) correctly describes 
the quantum numbers of the nucleus as $J^{P}=(\frac{3}{2})^{-}$ and
 predicts the correct value of the magnetic moment of $^7$Li. 
In addition to the spin-angular dependence carried by the wave function
 (\ref{wf32text})  we have assumed a simple $|\vec{r}|$ dependence of the
 wave function (\ref{wf32text})
\begin{equation}
|\Psi_{Li-7}^{3/2}|^2 \propto {\rm exp}(-\frac{3}{2}\frac{r^2}{R^2})
\label{density}
\end{equation}
with $R^2=5.7$ fm$^2$, which correctly reproduces the e.m. form factor of $^7$Li. 

In order to calculate the scattering cross section we form 
the usual Glauber series for the nuclear profile function 
and integrate it with the square of the nuclear wave function
 over positions of the nucleons. 
We have neglected effects of Fermi motion of nucleons for shadowing since
they are very small.
For the details of the formalism see Appendix B.

 We retain  the first, second and third terms
 of the expansion, which corresponds to single, double and triple scattering of a projectile off the target. The contribution of the quadrupole term is negligible since already  the triple scattering term gives only a less than $2\%$ contribution.
Since it is a fairly lengthy calculation, we refer the reader to Appendices B and C for details of our calculations of the total cross section of the polarized effective hadronic state $|h_{i} \rangle$ with helicity +
off polarized $^7$Li with $M_{J}=3/2$ and with $M_{J}=-3/2$, which we named 
$\sigma_{A}^{+,3/2}$ and $\sigma_{A}^{+,-3/2}$.

Next, using Appendices B and C,  we find  the difference of the cross sections   
with $M_{J}=3/2$ and  $M_{J}=-3/2$
\begin{eqnarray}
\Delta \sigma_{A}^{Li,3/2}&=&\sigma_{A}^{+,3/2}-\sigma_{A}^{+,-3/2}= \nonumber\\
&&\frac{13}{15}\Delta \sigma_{p}+\frac{2}{15}\Delta \sigma_{n}-\frac{\sigma_{eff}}{\pi(R^2+3B)}\alpha_{1}\Big(\frac{13}{15}\Delta \sigma_{p}+\frac{2}{15}\Delta \sigma_{n}\Big)F(x) \nonumber\\
&-&\frac{2\sigma_{eff}}{\pi R^2}\Big((\frac{9}{15}\alpha_{4}+\frac{4}{15}\alpha_{6})\Delta \sigma_{p}+\frac{2}{15}\alpha_{6}\Delta \sigma_{n}\Big)F(x)+\Big(\Delta \sigma_{p} \cdot 0.0143+\Delta \sigma_{n} \cdot 0.0025 \Big)g(x) \ .
\label{delta}
\end{eqnarray}
Here $\Delta \sigma_{p}=\sigma_{p}^{++}-\sigma_{p}^{+-}$ and $\Delta \sigma_{n}=\sigma_{n}^{++}-\sigma_{n}^{+-}$ are the differences of the $|h \rangle$-nucleon cross sections with parallel and anti-parallel helicities for protons and neutrons respectively.   
$F(x)={\rm exp}(-(q_{\parallel}R)^2/3)={\rm exp}(-176 \cdot x^2)$ originates from  the non-vanishing $q_{\parallel}$.
 $g(x)$ has the same origin. It is defined $g(0)$=1. Since it is a  function of $x$ slower than $F(x)$, in our numerical analysis we set $g(x)$=1 for any $x$ without any loss of accuracy of our results.
 $B$=6 GeV$^{-2}$ is the slope of the
$\left|h\right>$-nucleon elastic cross section. 
$\alpha_{1}=1.376$, $\alpha_{4}$=0.300, $\alpha_{6}$=0.402.
Since the triple scattering term has a complicated analytical structure, we simply give it in a numerical form (the last term in Eq.\ (\ref{delta})). 
Also note that the denominator of the forth term in the expression above is $1/R^2$ rather than $1/(R^2+3B)$. We have omitted $3B$ as compared to $R^2$ to get an analytical expression. It does not change our predictions or numerical results. We point out
 that 
$\Delta \sigma_{A}^{Li,3/2}$ arises due to orbital motion of the valence nucleons, which occupy the $1p_{3/2}$ shell.

  When the target is longitudinally polarized with $M_{J}$=1/2  we can find the difference of the cross sections of the polarized effective  state $|h \rangle$ off polarized $^7$Li  with $M_{J}$=1/2 and $M_{J}$=-1/2. Using Appendix D we obtain
\begin{eqnarray}
&&\Delta \sigma_{A}^{Li,1/2}=\sigma_{A}^{+,1/2}-\sigma_{A}^{+,-1/2}= \nonumber\\
&&\frac{1}{3}\Bigg(\frac{13}{15}\Delta \sigma_{p}+\frac{2}{15}\Delta \sigma_{n}-\frac{\sigma_{eff}}{\pi(R^2+3B)}\tilde{\alpha_{1}}\Big(\frac{13}{15}\Delta \sigma_{p}+\frac{2}{15}\Delta \sigma_{n}\Big)F(x) \nonumber\\
&-&\frac{2\sigma_{eff}}{\pi R^2}\Big((\frac{9}{15}\tilde{\alpha_{4}}+\frac{4}{15}\tilde{\alpha_{6}})\Delta \sigma_{p}+\frac{2}{15}\tilde{\alpha_{6}}\Delta \sigma_{n}\Big)F(x)+\Big(\Delta \sigma_{p} \cdot 0.0139+\Delta \sigma_{n} \cdot 0.0015 \Big)g(x) \Bigg) \ .
\label{delta12}
\end{eqnarray}
Here $\tilde{\alpha_{1}}=4\alpha_{2}-\alpha_{1}$=2.633, $\tilde{\alpha_{4}}=(16\alpha_{5}-\alpha_{4})/3$=0.866, $\tilde{\alpha_{6}}=4\alpha_{5}-\alpha_{4}$=0.424.
From Eq.\ (\ref{delta12}) one can see that $\Delta \sigma_{A}^{Li,1/2}$ is suppressed by approximately a factor of three as compared to   $\Delta \sigma_{A}^{Li,3/2}$. Therefore, the measurement of $\Delta \sigma_{A}^{Li,1/2}$ and the extraction of the corresponding spin dependent structure function is not as feasible as that of $\Delta \sigma_{A}^{Li,3/2}$ and the corresponding spin dependent structure function. Hence, in our paper we will make predictions of the amount of shadowing and enhancement  only for polarized $^7$Li with $M_{J}$=3/2.

In deep inelastic scattering experiments one can measure the
differential cross section of polarized projectiles off the longitudinally polarized target. The difference of such cross sections off the target polarized along the beam and in the opposite direction is called polarization asymmetry and it  is proportional to the corresponding spin dependent structure functions.
On the other hand, the polarization asymmetry is proportional to   
the differences of the cross sections given by Eqs.\ (\ref{delta}) and (\ref{delta12}). Therefore, in the Bjorken limit one can write \cite{Jaffe}
\begin{eqnarray}
\Delta \sigma_{A}^{Li,3/2}&\propto &\frac{d \sigma_{A}^{+,3/2}}{dx\,dy}-\frac{d \sigma_{A}^{+,-3/2}}{dx\,dy}=\frac{e^4M_{A}E}{2\pi Q^{4}}y(2-y)xg_{1}^{3/2\,3/2}(x) \nonumber\\
\Delta \sigma_{A}^{Li,1/2}&\propto &\frac{d \sigma_{A}^{+,1/2}}{dx\,dy}-\frac{d \sigma_{A}^{1,-3/2}}{dx\,dy}=\frac{e^4M_{A}E}{2\pi Q^{4}}y(2-y)xg_{1}^{1/2\,1/2}(x) \ .
\label{link}
\end{eqnarray}
Here $y=\nu/(M_{A}E)$, $x=Q^2/2\nu$, and $\nu=q \cdot P_{A}$.
Since spin of $^7$Li is 3/2, one has to introduce two different spin dependent structure functions $g_{1}^{3/2\,3/2}(x)$ and $g_{1}^{1/2\,1/2}(x)$. Note that the structure functions $g_{2}^{3/2\,3/2}(x)$ and $g_{2}^{1/2\,1/2}(x)$ are of higher twist and vanish in the Bjorken limit. In the quark parton language the spin dependent structure functions $g_{1}^{3/2\,3/2}(x)$ and $g_{1}^{1/2\,1/2}(x)$  are defined  \cite{Jaffe}
\begin{eqnarray}
&&g_{1}^{3/2\,3/2}(x) \equiv \frac{1}{2}\big(q_{\uparrow}^{3/2\,3/2}(x)-q_{\downarrow}^{3/2\,3/2}(x)\big) \nonumber\\
&&g_{1}^{1/2\,1/2}(x) \equiv \frac{1}{2}\big(q_{\uparrow}^{3/2\,1/2}(x)-q_{\downarrow}^{3/2\,1/2}(x)\big) \ ,
\end{eqnarray}
where $q_{\uparrow}^{3/2\,M_{J}}(x)$ ($q_{\downarrow}^{3/2\,M_{J}}(x))$ is defined to be the probability to find a quark with momentum fraction $x$ and spin up (down) in a target with spin-component $M_{J}$. The sum over quark flavors is assumed.

We make our predictions of the amount of shadowing and enhancement for the non-singlet combinations of the spin dependent structure functions defined
\begin{eqnarray} 
g_{1A=7}^{n.s.\,3/2}(x,Q^2) &\equiv& g_{1}^{^{7}{\rm Li}\,3/2\,3/2}(x,Q^2)-g_{1}^{^{7}{\rm Be}\,3/2\,3/2}(x,Q^2) \ , \nonumber\\ 
g_{1A=7}^{n.s.\,1/2}(x,Q^2) &\equiv& g_{1}^{^{7}{\rm Li}\,3/2\,1/2}(x,Q^2)-g_{1}^{^{7}{\rm Be}\,3/2\,1/2}(x,Q^2) \ , \nonumber\\ 
g_{1A=1}^{n.s.}(x,Q^2) &\equiv& g_{1}^{p}(x,Q^2)-g_{1}^{n}(x,Q^2) \ .
\end{eqnarray}

These non-singlet combinations for $A$=7 are proportional to $\Delta \sigma_{A}^{Li,3/2}-\Delta \sigma_{A}^{Be,3/2}$ ($\Delta \sigma_{A}^{Li,1/2}-\Delta \sigma_{A}^{Be,1/2}$)
by virtue of Eq.\ (\ref{link}).

 In order to get $\Delta \sigma_{A}^{Be,3/2}$ ($\Delta \sigma_{A}^{Be,1/2}$) from $\Delta \sigma_{A}^{Li,3/2}$ ($\Delta \sigma_{A}^{Li,1/2}$) one should simply substitute the valence proton by the valence neutron.  Therefore, using Eq.\ (\ref{delta})  
we  obtain for the ratio of the non-singlet structure functions with $A$=7 and $A$=1 for the nuclear target with $M_{J}$=3/2
\begin{eqnarray}
&&\frac{g_{1A=7}^{n.s.\,3/2}(x,Q^2)}{g_{1N}^{n.s.}(x,Q^2)}=\frac{\Delta \sigma_{A}^{Li,3/2}-\Delta \sigma_{A}^{Be,3/2}}{\Delta \sigma_{p}-\Delta \sigma_{n}} \nonumber\\
&=&\frac{11}{15}\times \Bigg(1- \frac{\sigma_{eff}}{\pi(R^2+3B)}\alpha_{1}\,F(x)-\frac{2\sigma_{eff}}{\pi R^2}(\frac{9}{15}\alpha_{4}+\frac{2}{15}\alpha_{6})\frac{15}{11}\,F(x)+0.0161\,g(x) \Bigg) \ .
\label{mainth}
\end{eqnarray} 

Substituting  $R^2=5.7$ fm$^2$, $\sigma_{eff}$=17 mbarn, $B=6$ GeV$^{-2}$ and numerical values of $\alpha$'s we obtain 
\begin{equation}
\frac{g_{1A=7}^{n.s.\,3/2}(x,Q^2)}{g_{1A=1}^{n.s}(x,Q^2)}=\frac{11}{15}\times \Big(1-0.177\,{\rm exp}(-176\,x^2)+0.016\,g(x)\Big) \ .
\label{main}
\end{equation}

This equation  is our main result for deep inelastic scattering on polarized $^7$Li. It predicts $16\%$ shadowing of $\frac{15}{11} \times g_{1A=7}^{n.s.\,3/2}(x,Q^2) / g_{1A=1}^{n.s.}(x,Q^2)$  for $x \leq 10^{-2}$ (see the solid line in Fig.\ (\ref{fig1})).

When the target is polarized longitudinally with $M_{J}$=1/2, we obtain for the ratio of the corresponding non-singlet structure function with $A$=7 and $A$=1
\begin{eqnarray}
&&\frac{g_{1A=7}^{n.s.\,1/2}(x,Q^2)}{g_{1N}^{n.s.}(x,Q^2)}=\frac{\Delta \sigma_{A}^{Li,1/2}-\Delta \sigma_{A}^{Be,1/2}}{\Delta \sigma_{p}-\Delta \sigma_{n}} \nonumber\\
&=&\frac{1}{3} \cdot \frac{11}{15}\times \Bigg(1- \frac{\sigma_{eff}}{\pi(R^2+3B)}\tilde{\alpha_{1}}\,F(x)-\frac{2\sigma_{eff}}{\pi R^2}(\frac{9}{15}\tilde{\alpha_{4}}+\frac{2}{15}\tilde{\alpha_{6}})\frac{15}{11}\,F(x)+0.0372\,g(x) \Bigg) \ .
\label{mainth12}
\end{eqnarray}
Numerically Eq.\ (\ref{mainth12}) gives
\begin{equation}
\frac{g_{1A=7}^{n.s.\,1/2}(x,Q^2)}{g_{1N}^{n.s.}(x,Q^2)}=\frac{1}{3} \cdot \frac{11}{15}\Big(1-0.372\,{\rm exp}(-176\,x^2)+0.037\,g(x)\Big) \ .
\label{main12}
\end{equation}
Comparing Eq.\ (\ref{main12}) with Eq.\ (\ref{main}) one can see that the structure function  $g_{1A=7}^{n.s.\,1/2}(x,Q^2)$ is suppressed by approximately  a factor of 3.5 in the region $x < 0.05$  as compared to 
$g_{1A=7}^{n.s.\,3/2}(x,Q^2)$. This makes it difficult to determine  this structure function in experiment.

Now we can prove our observation that this amount of shadowing for the polarized
structure functions 
is larger by approximately a factor of two than that for the unpolarized one.
In order to do so we first need to find the total scattering cross section of the unpolarized effective hadronic projectile $|h \rangle$ off the polarized  $^7$Li target with $M_{J}$=3/2 and $M_{J}$=1/2. We refer the reader to Appendix E for the details of this calculation.  We define the scattering cross section of the unpolarized $|h\rangle$ off the unpolarized $^7$Li target by
\begin{equation}
\sigma_{A}=\frac{1}{2}\Big(\sigma_{A}^{3/2}+\sigma_{A}^{1/2}\Big) \ .
\end{equation}
Here $\sigma_{A}^{3/2}$ and $\sigma_{A}^{1/2}$ are the scattering cross sections of the unpolarized $|h \rangle$ off the $^7$Li target with $M_{J}$=3/2 and $M_{J}$=1/2 respectively.
Then, using the results of Appendix E, we obtain for the ratio of  scattering cross sections of the unpolarized effective state $|h\rangle$ off unpolarized $^7$Li to $A\,\sigma_{eff}$ 
\begin{equation}
\frac{\sigma_{A}}{7\sigma_{eff}}=1-0.0996\,F(x)+0.0072\,g(x) \ .
\label{shunplrz}
\end{equation}
Eq.\ (\ref{shunplrz}) gives 9.2\% shadowing at $x \le$ 0.01.
Therefore, the amount of shadowing in the polarized case is  larger by a factor of two than that for the unpolarized case with a high accuracy.  

Semi-quantitatively one can see the origin of this factor of two from the following idealized argument.
 Let us assume that in $^7$Li and $^3$He all nuclear spin
 is due to the spin of a single valence nucleon. Then the Glauber expansion for the polarized target  reads
\begin{equation}
\frac{\Delta \sigma_{A}}{\Delta \sigma_{N}}=1-(A-1)\frac{\sigma_{eff}}{R^2}\,k +\dots  \ .
\label{shpol}
\end{equation}
Here  $R^2$ is the radius of the nucleus, $k$ is some numerical factor.   
Note that in Eq.\ (\ref{shpol}) we have kept only the first and the second terms since the higher terms are small.   
The factor $A-1$ originates from $A-1$ ways to couple the 
valence nucleon with the other nucleons. 
Eq.\ (\ref{shpol}) should be compared to 
\begin{equation}
\frac{\sigma_{A}}{A\sigma_{N}}=1-\frac{A-1}{2}\frac{\sigma_{eff}}{R^2}\,k +\dots \ ,
\label{shunpol}
\end{equation}
which describes shadowing in the case of the unpolarized target.
 Here $\sigma_{A}$ is the total spin-averaged $h \rangle$-nuclear cross section.
\mbox{$(A-1)/2$} comes from $A(A-1)/2$ ways to couple any two nucleons. 
One can see that in Eq.\ (\ref{shunpol}) the coefficient in front of 
the double scattering term is twice smaller than that in Eq.\ (\ref{shpol}).  
Thus, in the idealized situation of one valence nucleon our 
observation of the enhancement of shadowing by a factor of
 two for the polarized target is a consequence of a simple counting of pairs. 

For $^3$He this idealization of one valence nucleon is a very good approximation since almost all nuclear spin is carried by the neutron.
Consequently, in our calculations we used the wave function of $^3$He, where only the neutron is polarized.

It is interesting to compare the ratio of the non-singlet structure functions with $A$=7 and $M_{J}$=3/2 and $A$=1, given by Eq.\ (\ref{mainth}), and the ratio of the spin dependent structure functions $g_{1}^{3/2\,3/2}(x,Q^2)$ for $^7$Li and  a proton. The latter can found immediately from Eq.\ (\ref{delta})
\begin{eqnarray}
&&\frac{g_{1}^{{\rm Li}\,3/2\,3/2}(x,Q^2)}{g_{1}^{p}(x,Q^2)}=\frac{\Delta \sigma_{A}^{Li}}{\Delta \sigma_{p}}= \nonumber\\
&&\frac{13}{15}+\frac{2}{15}\frac{\Delta \sigma_{n}}{\Delta \sigma_{p}}-\frac{\sigma_{eff}}{\pi(R^2+3B)}\alpha_{1}\Big(\frac{13}{15}+\frac{2}{15}\frac{\Delta \sigma_{n}}{\Delta \sigma_{p}}\Big)F(x) \nonumber\\
&-&\frac{2\sigma_{eff}}{\pi R^2}\Big((\frac{9}{15}\alpha_{4}+\frac{4}{15}\alpha_{6})+\frac{2}{15}\alpha_{6}\frac{\Delta \sigma_{n}}{\Delta \sigma_{p}}\Big)F(x)+\Big(0.0143+\frac{\Delta \sigma_{n}}{\Delta \sigma_{p}}\cdot 0.0025 \Big)\,g(x) \ .
\label{side}
\end{eqnarray}

Although experimentally at $x \leq 0.05$ $g_{1p}(x,Q^2)$ is close to $-g_{1n}(x,Q^2)$, or $\Delta \sigma(p)$ is close to $-\Delta \sigma(n)$, we find the effect of this deviation to be still sizable.
If we denote this deviation by some function $M(x)$ defined by
\begin{equation}
\frac{\Delta \sigma_{n}}{\Delta \sigma_{p}}=-1-M(x) \ ,
\end{equation}
then we can present Eq.\ (\ref{side}) in the form
\begin{equation}
\frac{g_{1}^{{\rm Li}\,3/2\,3/2}(x,Q^2)}{g_{1}^{p}(x,Q^2)}=\frac{g_{1A=7}^{n.s.\,3/2}(x,Q^2)}{g_{1N}^{n.s.}(x,Q^2)}-M(x)\Big(0.1333-0.0257\,F(x)+0.0025\,g(x)\Big) \ .
\end{equation}
The first term is the ratio of non-singlet spin dependent structure functions presented by Eqs.\ (\ref{mainth}) and (\ref{main}).
 Shadowing for the ratio $g_{1}^{{\rm Li}\,3/2\,3/2}(x,Q^2)/g_{1}^{p}(x,Q^2)$ is presented by a curved dotted line in Fig.\ (\ref{fig2}). We used the tabulation of the spin dependent structure functions $g_{1A=1}^{n.s.}(x,Q_{0}^2)$, $g_{1}^{p}(x,Q_{0}^2)$ and $g_{1}^{n}(x,Q_{0}^2)$  at $Q_{0}^2=10$ GeV$^2$  from \cite{adeva} based on the most recent analysis of the SMC data.

 The presence of nuclear shadowing  in the spin dependent structure function  $g_{1}^{^{7}{\rm Li}\,3/2\,3/2}(x,Q^2)$ violates the relation, which follows from  the Bjorken sum rule \cite{FGS}

\begin{equation}
R=\frac{\int^{1}_{0}[g_{1}^{^{7}{\rm Li}\,3/2\,3/2}(x,Q^2)-g_{1}^{^{7}{\rm Be}\,3/2\,3/2}(x,Q^2)]dx}
{\int^{1}_{0}[g_{1}^{p}(x,Q^2)
- g_{1}^{n}(x,Q^2)]dx}={g_{A}(A=7)\over g_{A}(A=1)} \ ,
\label{Bj}
\end{equation}
where $g_{A}$ is the axial coupling constant for $\beta$ decay.

Note that the main difference between $g_{A}$($A$=7) and $g_{A}$($A$=1), i.e. between $g_{1A=7}^{n.s.\,3/2}(x,Q^2)$ and $g_{1A=1}^{n.s.}(x,Q^2)$, is caused by the orbital motion of valence nucleons, which occupy the $1p_{3/2}$ shell. This gives the factor 11/15=0.73. Nuclear shadowing at small $x$ decreases this value by additional $16\%$.
Addition quenching of $R$ by a factor of $\eta_{A}=0.91$ \cite{quench} 
 is caused by 
higher partial waves in the $^7$Li ground state-wave function as well as 
admixtures of 
non-nucleonic degrees of freedom.  

We suggest to model the $x\sim 0.1$ 
enhancement so that its contribution to $R$ compensates shadowing
effect in the Bjorken sum rule.
 We require that

\begin{itemize}

\item{Enhancement does not affect the region $x \leq 0.05$, where shadowing is saturated.}
\item{Enhancement is concentrated  at 
$0.2 \geq x \ge 0.05 (0.1)$
 and compensates shadowing at $x \approx 0.1$.
\begin{equation} 
\frac{15}{11}\int _{0}^{.2}dx \Big( g_{1}^{^{7}{\rm Li}\,3/2\,3/2}(x,Q^2)-g_{1}^{^{7}{\rm Be}\,3/2\,3/2}(x,Q^2)\Big) =\int _{0}^{.2}dx \Big( g_{1}^{p}(x,Q^2)-g_{1}^{n}(x,Q^2) \Big) \ .
\label{modeling}
\end{equation}
}
\end{itemize}

The shape of the curve, which describes the 
enhancement region, 
deserves a special discussion. We model enhancement according to our 
expectations suggested by  experimental data on the $x$ dependence of the 
EMC effect. While we know that shadowing extends to $x \approx 0.05$, 
it is not known, where is the cross over point from shadowing to enhancement.
 While  equality\ (\ref{modeling}) fixes the integrated contribution of 
enhancement, we can only guess how enhancement is distributed along $x$,
 or where it reaches the maximum. Since the main contribution to the 
nuclear shadowing in Eq.\ (\ref{modeling}) comes from the region
 $x \leq 0.03 \div 0.05$, the variation of the cross over point between 
$x=0.05$ and $x=0.1$ does not change significantly the contribution of 
shadowing to integral (\ref{modeling}). It only governs the spread of 
 enhancement in $x$ and its hight.

We model  enhancement  at normalization point $Q^2$=4 GeV$^2$ according to 
Eq.\ (\ref{modeling}).
To obtain it at larger $Q^2$ one has to use the QCD evolution of the spin 
dependent parton densities.  
 However, since the scaling violation between $Q^2$=4 GeV$^2$ and $Q^2$=10 GeV$^2$ is small
 we used the SMC  parameterization of the non-singlet  structure function 
$g_{1A=1}^{n.s}(x,Q_{0}^2)$ from \cite{adeva} at $Q^2$=10 GeV$^2$.
Hence, we discuss here only the predictions for $g_{1A=7}^{n.s\,3/2}(x,Q_{0}^2)$
 at $Q^2$=4-10 GeV$^2$.

Note also that although the  parameterization of the parton densities of \cite{adeva} 
 is limited by $x=0.003$, 
the contribution to the integral from the region $x < 0.003$  is very small. 
Therefore, in our numerical modeling of enhancement in
 integral\ (\ref{modeling}) the lower limit of integration was $x=0.003$.

Our results are presented in Figs.\ ({\ref{fig1}) and (\ref{fig2}).
The solid lines in Fig.\ ({\ref{fig1}) represents three possible scenarios of  \mbox{$15/11\cdot g_{1A=7}^{n.s.\,3/2}(x,Q^2)/g_{1A=1}^{n.s.}(x,Q^2)$} with the cross over points $x$=0.05, 0.075 and 0.1 as functions of $x$.
 Enhancement is modeled so that both nuclear shadowing and enhancement do not alter $R$. 
 The choice of the cross over point changes the peak value of enhancement considerably. We obtained 55\% shadowing at $x$=0.15 for the cross over point $x=0.1$, 42\% shadowing at $x$=0.138 for the cross over point $x=0.075$ and  20\% shadowing at $x$=0.125 for the cross over point $x=0.05$.

In Fig.\ (\ref{fig2}) we present one of the curves of \mbox{$15/11\cdot g_{1A=7}^{n.s.\,3/2}(x,Q^2)/g_{1A=1}^{n.s.}(x,Q^2)$} from  Fig.\ ({\ref{fig1}). The curved dotted line is our calculation of shadowing for the ratio  
$15/11 \cdot g_{1}^{^{7}{\rm Li}\,3/2\,3/2}(x,Q^2)/g_{1}^{p}(x,Q^2)$. The difference between the solid and dotted lines  illustrated that at low $x$ shadowing is different for $g_{1A=7}^{n.s.\,3/2}(x,Q^2)/g_{1A=1}^{n.s.}(x,Q^2)$ and $g_{1}^{^{7}{\rm Li}\,3/2\,3/2}(x,Q^2)/g_{1}^{p}(x,Q^2)$ due to a non-vanishing difference between $g_{1p}(x,Q^2)$ and $-g_{1n}(x,Q^2)$.

Note that in the both figures we have  
did not take into account the 
the quenching  factor  
$\eta_{A}=0.91$ determined from the $g_A$ measurements. 
In order to reflect this quenching the ratio 
$g_{1A=7}^{n.s.\,3/2}(x,Q^2)/g_{1A=1}^{n.s}(x,Q^2)$   
has to be 
multiplied by $\eta_{A}=0.91$.

From  Figs.\ (\ref{fig1}) and (\ref{fig2}) one can see that there are sizable nuclear effects at $10^{-4} \leq x \leq 0.2$ in the ratio of the spin dependent  structure functions $g_{1A=7}^{n.s.\,3/2}(x,Q^2)/g_{1A=1}^{n.s.}(x,Q^2)$.
 These effects have a  nontrivial $x$ dependence: $16\%
$ shadowing for $10^{-4} \leq x \leq 0.03$ and enhancement of an order of 20 (55)\% at $x \approx 0.125$ (0.15), if enhancement occupies the region  
$0.05 \leq x \leq 0.2$ ($0.1 \leq x \leq 0.2$).

\section{How many-nucleon description breaks down}
\label{manynucleon}

It is interesting that our analysis of the spin dependent structure
 function of $^7$Li  $g_{1}^{^{7}{\rm Li}\,3/2\,3/2}(x,Q^2)$ in a
 wide range of $0.2 > x > 10^{-3}$ shows how the many-nucleon description of
 $^7$Li becomes invalid. We shall show that one cannot describe deep inelastic
 scattering off polarized $^7$Li as scattering off a many-nucleon system. 
Therefore, one needs to introduce explicitly non-nucleonic degrees of
 freedom to explain the behavior  $g_{1}^{^{7}{\rm Li}\,3/2\,3/2}(x,Q^2)$. 
Our proof is based on the following observation.
 If the many-nucleon description breaks down, the following inequality holds  
\begin{equation}
g_{1}^{^{7}{\rm Li}\,3/2\,3/2}(x,Q^2) \neq a\,g_{1}^{p}(x,Q^2)+b\,g_{1}^{n}(x,Q^2) \ .
\label{ineq}
\end{equation}
Here $a$ and $b$ are some numerical factors,
which in
 the impulse approximation Eq.(\ref{impulse}) 
are equal to the polarizations of the valence proton and neutron.
They are also constrained by the Bjorken sum rule for $A=7$ system which leads to

\begin{equation}
{15\over 11}(a-b)=g_{A=7}/g_{A=1}=0.91 \ .
\end{equation}

 The factor $a$ is much larger than $b$ since the neutron contribution
 to the spin of $^7$Li is small.
 Although corrections to the nuclear shell model might 
change the coefficients $a$ and $b$, it will definitely
not be enough to turn Eq.\ (\ref{ineq}) into an equality.

 Since nuclear shadowing for the ratio 
$g_{1}^{^{7}{\rm Li}\,3/2\,3/2}(x,Q^2)/(13/15\,g_{1}^{p}(x,Q^2)+2/15\,g_{1}^{n}(x,Q^2))$
 is the same as for $15/11\,g_{1A=7}^{n.s.\,3/2}(x,Q^2)/g_{1A=1}^{n.s.}(x,Q^2)$ with
 a very hight accuracy, see Eq. (\ref{mainth}), we will make a realistic assumption
 that  enhancement is also equal for the both ratios.
Then if we rewrite Eq.\ (\ref{ineq}) in the form
\begin{equation}
\frac{g_{1}^{^{7}{\rm Li}\,3/2\,3/2}(x,Q^2)}{(13/15\,g_{1}^{p}(x,Q^2)+2/15\,g_{1}^{n}(x,Q^2))} \neq \frac{1}{g_{1}^{p}(x,Q^2)+2/11\,g_{1}^{D}(x,Q^2)}\big(a^{\prime}g_{1}^{p}(x,Q^2)+b^{\prime}g_{1}^{D}(x,Q^2)\big) \ ,
\label{ineq2}
\end{equation}
we can immediately compare its predictions with our predictions for enhancement given by the solid line in Figs.\ (\ref{fig1}) and (\ref{fig2}).
In Eq.\ (\ref{ineq2}) we used the deuteron spin dependent structure function $g_{1}^{D}(x,Q^2)=g_{1}^{p}(x,Q^2)+g_{1}^{n}(x,Q^2)$, $a^{\prime}=15/11(a-b)$ and $b^{\prime}=15/11b$. To obtain the best fit to nuclear shadowing by Eq.\ (\ref{ineq2}) we have chosen $a^{\prime}=0.90$ and $b^{\prime}=0.38$. 

In Fig.\ (\ref{fig3}) we plot the ratio $g_{1}^{^{7}{\rm Li}\,3/2\,3/2}(x,Q^2)/(13/15\,g_{1}^{p}(x,Q^2)+2/15\,g_{1}^{n}(x,Q^2))$ as a function of $x$.
The solid curved line is the same as in Fig.\ (\ref{fig2}).
 The dash-dotted line is given by Eq.\ (\ref{ineq2}). The
 clear inconsistency of both proves that inequality (\ref{ineq}) is valid,
 or that one cannot describe the discussed process with $^7$Li as with a
 many-nucleon system.
 If the many-nucleon description worked, $g_{1}^{^{7}{\rm Li}\,3/2\,3/2}(x,Q^2)$ 
could be approximated quite well by the right-hand-side of Eq.\ (\ref{ineq}). 
Due to the presence of  enhancement at 0.05 $(0.1) \leq x \leq 0.2$
 it is clearly impossible. 
 
Therefore, in order  to test the discussed ideas in experiment,
 one should explore the region of enhancement 0.05 $(0.1) \leq x \leq 0.2$, 
where the many-nucleon description of polarized deep inelastic scattering off 
$^7$Li with $M_{J}$=3/2  
is expected to break down completely.

\section{Differences in scattering of unpolarized leptons off $^3$He with $M_{J}$=3/2 and $M_{J}$=1/2.} 
\label{unpolarized}

As a by-product of our calculations we notice that  one can see  in experiment the difference in cross sections of deep inelastic scattering of  unpolarized leptons off $^7$Li polarized longitudinally  with $M_{J}$=3/2 and  $M_{J}$=1/2. Combining Eqs.\ (\ref{unpl32}) and (\ref{unpl12}) of Appendix E we obtain
\begin{equation}
\frac{\sigma_{A}^{3/2}-\sigma_{A}^{1/2}}{\sigma_{eff}}=-0.1029\,F(x)-0.0154\,g(x) \ ,
\label{ww2}
\end{equation}
which suggests a 12\% effect at $x \le 0.01$.
Although the quantity, which can be measured in experiment, is the ratio of $\sigma_{A}^{3/2}-\sigma_{A}^{1/2}$ to the unpolarized cross section $\sigma_{A}$.
Combining Eqs.\ (\ref{ww2}) and (\ref{shunplrz}) we find that at $x \le$ 0.01
\begin{equation}
\frac{\sigma_{A}^{3/2}-\sigma_{A}^{1/2}}{\sigma_{A}}=-0.0186 \ .
\end{equation}
Therefore, we predict a 2\% effect in the difference of scattering cross sections  of deep inelastic scattering of  unpolarized leptons off $^3$He with $M_{J}$=3/2 and $M_{J}$=1/2. This effect is due to the presence of higher partial waves in the wave function of $^7$Li. 
Note that a similar effect was pointed out for deuterium \cite{Piller,GW,Strikman,NS}.

\section{Nuclear effects at small $x$ in $g_{1A}(x,Q^2)$ for $^3$He}
\label{helium}

Another remarkable nucleus, which suits well for the purpose of studying spin dependent structure functions, is $^3$He. Nuclear polarization is  carried predominantly by a single nucleon, the neutron, which enhances nuclear effects of shadowing and enhancement in two times as compared to the unpolarized nucleus.
Below we will give a short account of our original paper \cite{FGS}, where 
more details and references can be found.
 
Using the modified Gribov-Glauber formalism we presented
 the total  cross section of the polarized hadronic state $|h\rangle$ with helicity +  on polarized $^3$He with helicity $\pm$ at $x \leq 0.05$  
\begin{eqnarray}
\sigma^{+,\pm}_{A}&=&\sigma^{+\,\pm}_{n}+2\sigma_{eff}-\frac{\sigma_{eff}^2e^{-\alpha\, q_{\parallel}^2}}{8\pi(\alpha+B)} \nonumber\\
&-&\frac{\sigma^{+\,\pm}_{n}\sigma_{eff}e^{-\alpha\, 
q_{\parallel}^2}}{4\pi(\alpha+B)}+\frac{\sigma_{eff}^2\sigma^{+,\pm}_{n}}{48\pi^2(\alpha+B)^2}\,f(x) \ ,
\label{cs1}  
\end{eqnarray}
where superscripts $++$ and $+-$ stand for parallel  and anti-parallel helicities of the incoming photon (the effective state $h$ has the same helicity) and the target nucleus or the neutron.
Here $\alpha=27$ GeV$^{-2}$ is the slope of a nuclear one-particle density chosen to reproduce the e.m. form factor of $^3$He. $B$=6 GeV$^{-2}$ is the slope of the $|h \rangle$-nucleon cross section. 
The function $f(x)$ is a function of $x$ slower than ${\rm exp}(-\alpha\, 
q_{\parallel}^2)$.

This leads to  the ratio of the spin dependent structure functions of $^3$He and a neutron \cite{FGS}
\begin{eqnarray}
\frac{g_{1}^{^{3}{\rm He}}(x,Q_0^2)}{g_{1}^{n}(x,Q_0^2)}&=&
\frac{\sigma_{T}^{+}(e^3He)-\sigma_{T}^{-}(e^3He)}
{\sigma_{T}^{+}(en)-\sigma_{T}^{-}(en)}=\nonumber\\
1&-&\frac{\sigma_{eff}e^{-\alpha\, q_{\parallel}^2}}{4\pi(\alpha+B)}+\frac{\sigma_{eff}^2}{48\pi^2(\alpha+B)^2}\,f(x) \ . 
\label{masterhe}
\end{eqnarray}

Numerically, for example at $x \leq 0.03$ $g_{1}^{^{3}{\rm He}}(x,Q_0^2) / g_{1^{n}}(x,Q_0^2)=0.9$, which provides the amount of nuclear shadowing, which  is by a factor of two larger than the corresponding amount for unpolarized structure functions, $F_{2A=3}(x,Q^2)/3F_{2N}(x,Q^2)=0.95$.

Exactly as in the case of $^{7}$Li 
we  model enhancement so that its contribution to $R$ compensates shadowing. 
We present our results in Figs.\ (\ref{fig4}) and (\ref{fig5}).
These figures are somewhat different from the original figure presented in 
\cite{FGS},  where 
the $x$ dependence of proton and neutron spin dependent structure functions
was not properly taken into account.
Now we used the most recent parameterization  from \cite{adeva}.

 We plot
\mbox{$g_{1A=3}^{n.s.}(x,Q^2)/g_{1N}^{n.s.}(x,Q^2)=
(g_{1}^{^{3}{\rm He}}(x,Q^2)-g_{1}^{^{3}{\rm H}}(x,Q^2))
/(g_{1}^{n}(x,Q^2)- g_{1}^{p}(x,Q^2))$} as a function of $x$.
Unlike $^7$Li, since we assumed, that only one nucleon carries all nuclear polarization, 
shadowing is the same for the ratios $g_{1}^{^{3}\rm{He}}(x,Q^2)/g_{1}^{n}(x,Q^2)$, $g_{1}^{^{3}\rm{H}}(x,Q^2)/g_{1}^{p}(x,Q^2)$ and $g_{1A=3}^{n.s.}(x,Q^2)/g_{1N}^{n.s.}(x,Q^2)$.
In Fig.\ (\ref{fig4}) we present three possible scenarios of enhancement, which depend on the cross over point, like in the case of $^7$Li. 

We obtained 40\% enhancement at $x$=0.15 for the cross over point $x=0.1$, 26\% enhancement at $x$=0.138 for the cross over point $x=0.075$ and  14\% enhancement at $x$=0.125 for the cross over point $x=0.05$.

In Fig.\ (\ref{fig5}) the dash-dotted straight line represents the ratio $R$ in the impulse approximation, which includes also higher partial waves ($S^{\prime}$, $D$, etc.) in the ground-state wave function of $^3$He. It corresponds to the $x$ independent ratio of the spin dependent structure functions  \mbox{$g_{1A=3}^{n.s.}(x,Q^2)/g_{1N}^{n.s.}(x,Q^2)$=0.9215}.
The dotted straight line represents $R$  within the impulse approximation corrected to include non-nucleonic degrees of freedom in the ground-state wave function of $^3$He. It corresponds to
\mbox{$g_{1A=3}^{n.s.}(x,Q^2)/g_{1N}^{n.s.}(x,Q^2)$=0.9634}.
The curved dotted line is a result of our calculations of nuclear shadowing and  modeling of enhancement. 
We assume that the discussed effects contribute multiplicatively, which shifts the  curved dotted line  downward.  Our predictions for $g_{1A=3}^{n.s.}(x,Q^2)/g_{1N}^{n.s.}(x,Q^2)$, $g_{1}^{^{3}\rm{He}}(x,Q^2)/g_{1}^{n}(x,Q^2)$ and $g_{1}^{^{3}\rm{H}}(x,Q^2)/g_{1}^{p}(x,Q^2)$  are given by the solid line.

 We conclude that similar to $^7$Li, nuclear effects at $10^{-4} \leq x \leq 0.2$ in the ratio of the  spin dependent nuclear structure functions $g_{1A=3}^{n.s.}(x,Q^2)/g_{1N}^{n.s.}(x,Q^2)$ are large:  shadowing is of an order of $10\%$ for $10^{-4} \leq x \leq 0.03$ and enhancement is of an order of 
14 (40)\% at  $x \approx 0.125$ (0.15), if  enhancement occupies the region $0.05 \leq x \leq 0.2$ ($0.1 \leq x \leq 0.2$).

\section{Nuclear shadowing for polarized gluons}

As we pointed out in Sec.\ ref{lithium} the nuclear shadowing in the gluon channel is larger by a factor of three than that in the quark channel. 
Although the contribution of the gluon channel to the polarized DIS has not received a solid experimental ground, we still can make prediction for the amount of nuclear shadowing for the polarized nuclear gluon parton density $\Delta G_{A}(x,Q^2)$.
Similarly to Eq.\ (\ref{masterhe}) one can estimate the amount of nuclear shadowing for $\Delta G_{A}(x,Q^2)$, see also \cite{FS981}
\begin{equation}
\frac{\Delta G_{A}(x,Q^2)}{\Delta G_{N}(x,Q^2)}=
1-\frac{\sigma_{eff}e^{-\alpha\, q_{\parallel}^2}}{4\pi(\alpha+B)}+\frac{\sigma_{eff}^2}{48\pi^2(\alpha+B)^2}\,f(x) \ . 
\label{masterglue}
\end{equation}
Here $\Delta G_{N}(x,Q^2)$ is the the polarized nucleon gluon parton density.
Using $\sigma_{eff}$=50 mbarn in Eq.\ (\ref{masterglue}) we will find at $x \leq 0.03$
\begin{equation}
\frac{\Delta G_{A}(x,Q^2)}{\Delta G_{N}(x,Q^2)}=0.70 \ .
\end{equation}
This means that the effect of nuclear shadowing for polarized gluons is of an order of 30\%. In the future it may be possible to measure this effect at HERA when polarized protons are used.

\section{Conclusions}
\label{conclusion}

We propose to use polarized nuclear targets of $^7$Li and  $^3$He for studying  nuclear effects in the spin dependent structure functions $g_{1A}(x,Q^2)$, where these effects are expected to be enhanced by a factor of two as compared to the unpolarized targets.
 We predict a significant $x$ dependence at  $10^{-4} \leq x \leq 0.2$ due to the effects of nuclear shadowing and enhancement. The effect of nuclear shadowing is of an order of $16\%$ for the ratio $g_{1A=7}^{n.s.}(x,Q^2)/g_{1N}^{n.s.}(x,Q^2)$  and  $10\%$ for $g_{1A=3}^{n.s.}(x,Q^2)/g_{1N}^{n.s.}(x,Q^2)$. By imposing the requirement that
 the Bjorken sum rule is satisfied we model the effect of enhancement. We find the effect of enhancement at $x \approx 0.125$ (0.15) to be of an order of (20) 55\% for $g_{1A=7}^{n.s.}(x,Q^2)/g_{1N}^{n.s.}(x,Q^2)$  and 14 (40)\% for $g_{1A=3}^{n.s.}(x,Q^2)/g_{1N}^{n.s.}(x,Q^2)$ . We also point out that since the discussed  nuclear  effects in $^3$He are quite sizable, one should take them into 
account when extracting the neutron spin dependent structure function from $^3$He data.

We predict even larger nuclear shadowing effect for the polarized  
gluon densities. 

 We predict   a 2\% effect in the difference of scattering cross sections  of deep inelastic scattering of  unpolarized leptons off $^3$He longitudinally polarized with $M_{J}$=3/2 and $M_{J}$=1/2.
As in the deuteron, this effect is attributed to the presence of higher 
partial waves in the wave function of $^7$Li.

 We also predict a gross deviation from   the many-nucleon description of deep inelastic scattering off $^7$Li. The effect is pronounced in the enhancement region $0.2 > x > 0.05$, where we suggest to study it experimentally.

\acknowledgments{One of us (M.S.) would like to thank DESY
for the hospitality during the time this work was done.
We thank  D.~Crabb, L.~Frankfurt, J.~Lichtenstadt and   W.~Weise 
for useful discussions. This work is supported in part by the U.S. Department of
Energy.}

\newpage
\appendix
\section{The ground-state wave function of $^7$Li with $M_{J}$=3/2, 1/2, -1/2, -3/2.}
\label{appendix A}

The spin of $^7$Li is 3/2. In the shell model the ground-state wave function of $^7$Li with the $z$ component of the total angular momentum $M_{J}$=3/2 is given by \cite{Landau}
\begin{eqnarray}
\Psi_{Li-7}^{3/2}&=&\frac{3}{\sqrt{15}}[\Psi_{p}^{3/2}\Psi_{n}^{3/2}\Psi_{n}^{-3/2}]-\frac{2}{\sqrt{15}}[\Psi_{p}^{3/2}\Psi_{n}^{1/2}\Psi_{n}^{-1/2}] \nonumber\\
&-&\frac{1}{\sqrt{15}}[\Psi_{p}^{1/2}\Psi_{n}^{3/2}\Psi_{n}^{-1/2}]+\frac{1}{\sqrt{15}}[\Psi_{p}^{-1/2}\Psi_{n}^{3/2}\Psi_{n}^{1/2}] \ .
\label{wf32}
\end{eqnarray}

In order to find the wave function of $^7$Li with $M_{J}=1/2$ we act upon the wave function with $M_{J}=3/2$ by the lowering operator $L_{-}$. Using the relations
\begin{eqnarray}
&&L_{-}\Psi^{3/2}=\sqrt 3 \Psi^{1/2} \ , \nonumber\\ 
&&L_{-}\Psi^{1/2}=2 \Psi^{-1/2} \ , \nonumber\\ 
&&L_{-}\Psi^{-1/2}=\sqrt 3 \Psi^{-3/2} \ , \nonumber\\ 
&&L_{-}\Psi^{-3/2}=0 \ ,
\end{eqnarray}
we obtain the nuclear wave function with $M_{J}=1/2$
\begin{eqnarray}
\Psi_{Li-7}^{1/2}&=&\frac{1}{\sqrt{15}}[\Psi_{p}^{3/2}\Psi_{n}^{1/2}\Psi_{n}^{-3/2}]+\frac{2}{\sqrt{15}}[\Psi_{p}^{1/2}\Psi_{n}^{3/2}\Psi_{n}^{-3/2}] \nonumber\\
&-&\frac{3}{\sqrt{15}}[\Psi_{p}^{1/2}\Psi_{n}^{1/2}\Psi_{n}^{-1/2}]+\frac{1}{\sqrt{15}}[\Psi_{p}^{-3/2}\Psi_{n}^{3/2}\Psi_{n}^{1/2}] \ .
\label{wf12}
\end{eqnarray}

In a similar way we obtain the ground state-wave function of $^7$Li with $M_{J}=-1/2$
\begin{eqnarray}
\Psi_{Li-7}^{-1/2}&=&\frac{1}{\sqrt{15}}[\Psi_{p}^{3/2}\Psi_{n}^{-1/2}\Psi_{n}^{-3/2}]+\frac{2}{\sqrt{15}}[\Psi_{p}^{-1/2}\Psi_{n}^{3/2}\Psi_{n}^{-3/2}] \nonumber\\
&-&\frac{3}{\sqrt{15}}[\Psi_{p}^{-1/2}\Psi_{n}^{1/2}\Psi_{n}^{-1/2}]+\frac{1}{\sqrt{15}}[\Psi_{p}^{-3/2}\Psi_{n}^{3/2}\Psi_{n}^{-1/2}] 
\label{wf-12}
\end{eqnarray}
and with $M_{J}=-3/2$
\begin{eqnarray}
\Psi_{Li-7}^{-3/2}&=&\frac{3}{\sqrt{15}}[\Psi_{p}^{-3/2}\Psi_{n}^{3/2}\Psi_{n}^{-3/2}]-\frac{2}{\sqrt{15}}[\Psi_{p}^{-3/2}\Psi_{n}^{1/2}\Psi_{n}^{-1/2}] \nonumber\\
&-&\frac{1}{\sqrt{15}}[\Psi_{p}^{-1/2}\Psi_{n}^{1/2}\Psi_{n}^{-3/2}]+\frac{1}{\sqrt{15}}[\Psi_{p}^{1/2}\Psi_{n}^{-1/2}\Psi_{n}^{-3/2}] \ .
\label{wf-32}
\end{eqnarray}
\newpage
\section{Scattering off polarized $^7$Li targets with $M_{J}$=3/2 and $M_{J}$=-3/2. Single and double scattering.}
\label{Appendix B}

In this appendix we will give detailed calculations of the total scattering cross section of a polarized incoming hadron with helicity + off a polarized target of $^7$Li with the $z$ component of the total angular momentum $M_{J}=3/2$ and $M_{J}=-3/2$. When $M_{J}=3/2$ the target is polarized along the beam polarization. If $M_{J}$=-3/2 the target is polarized in the direction opposite  to the beam polarization.

We used the modified Glauber-Gribov formalism. In this formalism the total nuclear cross section $\sigma_{A}$ is related to the nuclear profile function $\Gamma_{A}(\vec{b})$ as
\begin{equation}
\sigma_{A}=2\,Re \int d^{2} \vec{b}\, \Gamma_{A}(\vec{b}) \ .
\end{equation}
Here  $\vec{b}$ is the impact parameter of the incoming particle with respect to the center of mass of the target nucleus. 
The nuclear profile function can be expanded as a series over nucleon profile functions $\Gamma(\vec{b}-\vec{s_{i}})$, where $\vec{s_{i}}$ is the transverse position of an $i$-th nucleon.  In the series we will keep only single, double and triple terms in accordance with our observation that higher terms give a negligible contribution in our case
\begin{eqnarray}
\Gamma_{A}(\vec{b})&=&\langle \Psi_{Li-7}^{3/2}|\sum_{i=1}^{7}\Gamma(\vec{b}-\vec{s_{i}})+\sum_{i=1, j > i}^{7}\Gamma(\vec{b}-\vec{s_{i}})\Gamma(\vec{b}-\vec{s_{j}})\Theta(z_{j}-z_{i})e^{iq_{\parallel}(z_{i}-z_{j})} \nonumber\\
&-&\sum_{i=1, j > i, k > j}^{7}\Gamma(\vec{b}-\vec{s_{i}})\Gamma(\vec{b}-\vec{s_{j}})\Gamma(\vec{b}-\vec{s_{k}})\Theta(z_{j}-z_{i})\Theta(z_{k}-z_{j})e^{iq_{\parallel}(z_{i}-z_{k})}|\Psi_{Li-7}^{3/2} \rangle \ .
\label{gammaA}
\end{eqnarray} 
Note that  the series is averaged with the ground state-wave function $\Psi_{Li-7}^{3/2}$.
 
This formula is different from the usual Glauber series. The exponents account for a non-zero $q_{\parallel}$. The theta-functions fix time-ordering of elementary scattering processes. The numerical factors in front of the second and third terms are used because there are two ways to time-order a pair of nucleons and there are six ways to time-order three nucleons. 

 The nucleon profile function is related to the scattering amplitude $f(\vec{k_{t}})$
\begin{equation}
\Gamma(\vec{b}-\vec{s_{i}})=\frac{1}{2\pi i k_{t}} \int d^2 \vec{k_{t}}\,e^{i\vec{k_{t}}(\vec{b}-\vec{s_{i}})} f(\vec{k_{t}}) \ .
\label{gamma}
\end{equation}  
The scattering amplitude at hight energies is predominantly imaginary
\begin{equation}
f(\vec{k_{t}})=\frac{ik}{4\pi}\sigma e^{-B/2  k_{t}^2} \ .
\label{ampl}
\end{equation}
Here $B$ is the slope of the hadron-nucleon cross section, $\sigma$ is the hadron-nucleon cross section.

Combining Eqs.\ (\ref{gamma}) and (\ref{ampl}) we obtain the nucleon profile function as a function of the nucleon transverse coordinates $\vec{s_{i}}$
\begin{equation}
\Gamma(\vec{b}-\vec{s_{i}})=\frac{\sigma}{4\pi B}\,e^{-(\vec{b}-\vec{s_{i}})^2/2B} \ .
\end{equation}
We used this expression of the nucleon profile function in series\ (\ref{gammaA}). We have neglected effects of Fermi motion of nucleons in the nucleus since these effects are negligible at $x < 0.5$. Next we will deal with single, double and triple terms of this series.

\underline{Single scattering}

We are computing $\sigma_{A}^{+,3/2}$, the total scattering cross section of a hadronic projectile with helicity + off a target of $^7$Li with $M_{J}$=3/2. The contribution of four non-valence nucleons in the $1s_{1/2}$ shell  is simply  $4\sigma_{eff}$. The contribution of the three valence nucleons is governed by the wave function\ (\ref{wf32text}). Since each valence nucleon belongs to the $1p_{3/2}$ shell, its total angular momentum is 3/2. Then, the spin-angular wave function of a valence nucleon with the $z$ component of the total angular momentum $m_{j}$=3/2, 1/2, -1/2 or -3/2 is
\begin{eqnarray}
\Psi_{N}^{3/2}&=&Y_{11}(\theta,\phi)|\uparrow\rangle=-\sqrt{\frac{3}{8\pi}}sin\,\theta\, e^{i\phi}|\uparrow \rangle \ , \nonumber\\ 
\Psi_{N}^{1/2}&=&\frac{1}{\sqrt{3}}Y_{11}(\theta,\phi)|\downarrow \rangle+\sqrt{\frac{2}{3}}Y_{10}(\theta,\phi)|\uparrow \rangle= \nonumber\\
&-&\frac{1}{\sqrt{8\pi}}sin\,\theta\, e^{i\phi}|\downarrow \rangle+\frac{1}{\sqrt{2\pi}}cos\,\theta |\uparrow \rangle \ , \nonumber\\ 
\Psi_{N}^{-1/2}&=&\sqrt{\frac{2}{3}}Y_{10}(\theta,\phi)|\uparrow \rangle+\frac{1}{\sqrt{3}}Y_{1-1}(\theta,\phi) |\downarrow \rangle= \nonumber\\ 
&+&\frac{1}{\sqrt{2\pi}}cos\,\theta |\uparrow \rangle-\frac{1}{\sqrt{8\pi}}sin\, \theta\, e^{-i\phi}|\downarrow \rangle \ , \nonumber\\ 
\Psi_{N}^{-3/2}&=&Y_{1-1}(\theta,\phi)|\downarrow \rangle=\sqrt{\frac{3}{8\pi}}sin\,\theta\, e^{-i\phi}|\downarrow \rangle \ .
\label{list} 
\end{eqnarray}
Here the superscripts stand for $m_{J}$.
$\theta$ is the polar angle, $\phi$ is the azimuthal angle, $|\uparrow \rangle$ and $|\downarrow \rangle$ are the spin-up and spin-down  nucleon spin states.

From Eq.\ (\ref{list}) one can see  that a valence nucleon with $m_{J}=3/2$ can have only spin up, a valence nucleon with $m_{J}=1/2$ has a 67\% probability to have its spin up and a 33\% probability to have its spin down, a valence nucleon with $m_{J}=-1/2$ has a 33\% probability to have its spin up and a 67\% probability to have its spin down, and  a valence nucleon with $m_{J}=-3/2$ can have only spin down. 

Combining this observation with the wave function with $M_{J}$=3/2\ (\ref{wf32text}), we obtain the contribution of the three valence nucleons to the cross section
\begin{equation}
\sigma_{A}^{+,3/2}=\frac{13}{15}\sigma_{p}^{++}+\frac{2}{15}\sigma_{n}^{++}+2\,\sigma_{eff} \ .
\label{singleval}
\end{equation}
Here $\sigma_{p}^{++}$ ($\sigma_{n}^{++}$) is the $|h\rangle$-proton (nucleon) cross section with parallel spins. 
In order to obtain Eq.\ (\ref{singleval}) we used the equality
\begin{eqnarray}
\sigma_{p}^{++}+\sigma_{p}^{+-}&=&2\,\sigma_{eff} \ , \nonumber\\ 
\sigma_{n}^{++}+\sigma_{n}^{+-}&=&2\,\sigma_{eff} \ .
\end{eqnarray}
Therefore, the total single scattering contribution is
\begin{equation}
\sigma_{A}^{+,3/2}=\frac{13}{15}\sigma_{p}^{++}+\frac{2}{15}\sigma_{n}^{++}+6\,\sigma_{eff} \ .
\end{equation} 
Note that the last term does not contribute to  $\Delta \sigma_{A}^{Li}$.

\underline{Double scattering}

Double scattering terms can be of three origins. We can form a pair by coupling six times non-valence nucleons, by coupling four times a valence and a non-valence nucleons, and by coupling three times valence nucleons. 

The contribution of the non-valence-non-valence pairs can be computed analytically without numerical integration
\begin{equation}
\sigma_{A}^{+,3/2}=-\frac{9}{4}\frac{\sigma_{eff}^2}{\pi(R^2+3B)}\cdot F(x) \ ,
\label{dblnvl}
\end{equation}
where $F(x)={\rm exp}(-(q_{\parallel}R)^2/3)={\rm exp}(-176 \cdot x^2)$, which originates from  the non-vanishing $q_{\parallel}$.
Note that since Eq.\ (\ref{dblnvl})  is spin-independent, it does not contribute to $\Delta \sigma_{A}^{Li}$  .

The contribution of the valence-non-valence pairs requires already some numerical integration due to the presence of nontrivial dependence on coordinates of the wave function of the valence nucleons. It is
\begin{equation}
\sigma_{A}^{+,3/2}=-\frac{\sigma_{eff}}{\pi(R^2+3B)}\Big(\frac{13}{15}\alpha_{1}\sigma_{p}^{++}+\frac{2}{15}\alpha_{1}\sigma_{n}^{++}+\sigma_{eff}(\frac{18}{15}\alpha_{1}+\frac{12}{15}\alpha_{3})\Big)F(x) \ .
\end{equation}
Here $\alpha_{1}$=1.376 and $\alpha_{3}$=1.795. Although the $x$ dependence cannot be found in this case in an analytical form, it is very close to $F(x)$. That is why we use the same $F(x)$ for the whole double scattering term.
 
The contribution of the three valence-valence pairs is quite bulky partially due to the fact that we would like to present it in its natural form, which keeps track of the numerical factors in the wave function\ (\ref{wf32text})
\begin{eqnarray}
\sigma_{A}^{+,3/2}&=&-\frac{2\sigma_{eff}}{\pi R^2}\Big(\frac{9}{15}\alpha_{4}\sigma_{p}^{++}+\frac{4}{15}\alpha_{6}\sigma_{p}^{++}+\frac{2}{15}\alpha_{6}\sigma_{n}^{++}\Big)F(x) \nonumber\\
&-&\frac{1}{\pi R^2}\Big(\frac{9}{15}\alpha_{4}\sigma_{n}^{++}\sigma_{n}^{+-}+
\frac{1}{15}(\frac{1}{9}\alpha_{4}(\sigma_{p}^{+-}\sigma_{n}^{++}+\sigma_{p}^{++}\sigma_{n}^{+-})+\frac{8}{9}\alpha_{5}(\sigma_{p}^{++}\sigma_{n}^{++}+\sigma_{p}^{+-}\sigma_{n}^{+-}) \nonumber\\
&+&\frac{16}{9}\alpha_{5}(\sigma_{p}^{++}\sigma_{n}^{+-}+\sigma_{p}^{+-}\sigma_{n}^{++})\Big) \nonumber\\
&+&\frac{4}{15}(\frac{1}{9}\alpha_{4}\sigma_{n}^{++}\sigma_{n}^{+-}+\frac{4}{9}\alpha_{5}(\sigma_{n}^{++}\sigma_{n}^{++}+\sigma_{n}^{+-}\sigma_{n}^{+-})+\frac{16}{9}\alpha_{5}\sigma_{n}^{++}\sigma_{n}^{+-})\Big)F(x) \ .
\end{eqnarray}
Here $\alpha_{4}$=0.300, $\alpha_{5}$=0.181, $\alpha_{6}$=0.402.

Combining all three contributions we find the double scattering term of the total scattering cross section of the projectile with helicity + off $^7$Li with $M_{J}$=3/2
\begin{eqnarray} 
\sigma_{A}^{+,3/2}&=&-\frac{9}{4}\frac{\sigma_{eff}^2}{\pi(R^2+3B)}\cdot F(x) \nonumber\\
&-&\frac{\sigma_{eff}}{\pi(R^2+3B)}\Big(\frac{13}{15}\alpha_{1}\sigma_{p}^{++}+\frac{2}{15}\alpha_{1}\sigma_{n}^{++}+\sigma_{eff}(\frac{18}{15}\alpha_{1}+\frac{12}{15}\alpha_{3})\Big)F(x) \nonumber\\
&-&\frac{2\sigma_{eff}}{\pi R^2}\Big(\frac{9}{15}\alpha_{4}\sigma_{p}^{++}+\frac{4}{15}\alpha_{6}\sigma_{p}^{++}+\frac{2}{15}\alpha_{6}\sigma_{n}^{++}\Big)F(x) \nonumber\\
&-&\frac{1}{\pi R^2}\Big(\frac{9}{15}\alpha_{4}\sigma_{n}^{++}\sigma_{n}^{+-}+
\frac{1}{15}(\frac{1}{9}\alpha_{4}(\sigma_{p}^{+-}\sigma_{n}^{++}+\sigma_{p}^{++}\sigma_{n}^{+-})+\frac{8}{9}\alpha_{5}(\sigma_{p}^{++}\sigma_{n}^{++}+\sigma_{p}^{+-}\sigma_{n}^{+-}) \nonumber\\
&+&\frac{16}{9}\alpha_{5}(\sigma_{p}^{++}\sigma_{n}^{+-}+\sigma_{p}^{+-}\sigma_{n}^{++})\Big) \nonumber\\
&+&\frac{4}{15}(\frac{1}{9}\alpha_{4}\sigma_{n}^{++}\sigma_{n}^{+-}+\frac{4}{9}\alpha_{5}(\sigma_{n}^{++}\sigma_{n}^{++}+\sigma_{n}^{+-}\sigma_{n}^{+-})+\frac{16}{9}\alpha_{5}\sigma_{n}^{++}\sigma_{n}^{+-})\Big)F(x) \ .
\end{eqnarray}

In order to find the total scattering cross section on polarized $^7$Li with $M_{J}$=-3/2, $\sigma_{A}^{+,-3/2}$, one should simply switch plus and minus signs in the second place in all the formulas above.

\section{Scattering off polarized $^7$Li targets with $M_{J}$=3/2 and $M_{J}$=-3/2. Triple scattering.}
\label{Appendix C}

The contribution of the triple scattering term to $\Delta \sigma_{A}^{Li}$  is only 1.6\% of the single scattering term and to $\sigma_{A}$ is  only  0.7\% of the single scattering term. To compute its contribution we can combine three non-valence nucleons, two non-valence and one valence nucleon, two valence and one non-valence nucleon and three valence nucleons. Due to the smallness and complexity of the triple scattering contribution we will give it in a numerical form and only the part of it, which contributes to $\Delta \sigma_{A}^{Li}$
\begin{equation}
\sigma_{A}^{+,3/2}=\Big(0.0143\,\sigma_{p}^{++}+0.0025\,\sigma_{n}^{++}\Big) \ .
\label{ttr}
\end{equation}
Due to cross section fluctuations in the incoming photon, the scattering formalism requires that the triple scattering term should contain $\langle \sigma^3 \rangle$, where one averages over fluctuations of the size of the projectile. Our assumption that the incoming photon interacts with the target through the effective  state $|h \rangle$ means that we have replaced $\langle \sigma^3 \rangle$ by $(\langle \sigma^2 \rangle)^2/ \langle \sigma \rangle=\sigma_{eff}^2 \langle \sigma \rangle$. 

In Eq.\ (\ref{ttr}) $g(x)$ is a function of $x$ slower than $F(x)$, defined $g(0)$=1. In our numerical analysis we set $g(x)$=1 for any $x$ without any loss of accuracy of our results.

In order to find the triple scattering term on polarized $^7$Li with $M_{J}$=-3/2, $\sigma_{A}^{+,-3/2}$, one should simply switch plus and minus signs in  the formula above.
 
\newpage
\section{Scattering off polarized $^7$Li targets with $M_{J}$=1/2 and $M_{J}$=--1/2.}
\label{Appendix D}
 
In this appendix we will give detailed calculations of the total scattering cross section of a polarized incoming hadron with helicity + off a polarized target of $^7$Li with the $z$ component of the total angular momentum $M_{J}=1/2$ and $M_{J}=-1/2$. 
Following the steps of the Glauber-Gribov formalism described in Appendix B, we will present the single, double and triple scattering terms.  

\underline{Single scattering}

The total single scattering contribution is a sum of the valence contribution and a simple non-valence contribution $4\,\sigma_{eff}$
\begin{equation}
\sigma_{A}^{+,1/2}=\frac{13}{15}\Big(\frac{2}{3}\sigma_{p}^{++}+\frac{1}{3}\sigma_{p}^{+-}\Big)+\frac{2}{15}\Big(\frac{2}{3}\sigma_{n}^{++}+\frac{1}{3}\sigma_{n}^{+-}\Big)+6\,\sigma_{eff} \ . 
\end{equation} 

\underline{Double scattering}

As in the case of $M_{J}=3/2$, the double scattering term originates from the three possible ways to form  nucleon-nucleon pairs. Similarly to the calculation given in Appendix B we 
combine all three contributions in order to find the double scattering term of the total scattering cross section of the projectile with helicity + off $^7$Li with $M_{J}$=1/2
\begin{eqnarray} 
&&\sigma_{A}^{+,1/2}=-\frac{9}{4}\frac{\sigma_{eff}^2}{\pi(R^2+3B)}\cdot F(x) \nonumber\\
&-&\frac{\sigma_{eff}}{\pi(R^2+3B)}\Big(\frac{13}{15}(\frac{1}{3}\alpha_{1}\sigma_{p}^{+-}+\frac{4}{3}\alpha_{2}\sigma_{p}^{++})+\frac{2}{15}(\frac{1}{3}\alpha_{1}\sigma_{n}^{+-}+\frac{4}{3}\alpha_{2}\sigma_{n}^{++})
+\sigma_{eff}(\frac{12}{15}\alpha_{1}+\frac{18}{15}\alpha_{3})\Big)F(x) \nonumber\\
&-&\frac{2\sigma_{eff}}{\pi R^2}\Big(\frac{9}{15}(\frac{1}{9}\alpha_{4}\sigma_{p}^{+-}+\frac{8}{9}\alpha_{5}\sigma_{eff}+\frac{16}{9}\alpha_{5}\sigma_{p}^{++})+\frac{4}{15}(\frac{1}{3}\alpha_{4}\sigma_{p}^{+-}+\frac{4}{3}\alpha_{5}\sigma_{p}^{++}) \nonumber\\
&+&\frac{2}{15}(\frac{1}{3}\alpha_{4}\sigma_{n}^{+-}+\frac{4}{3}\alpha_{5}\sigma_{n}^{++})\Big)F(x) \nonumber\\
&-&\frac{1}{\pi R^2}\Big(\frac{9}{15}((\frac{1}{9}\alpha_{4}+\frac{16}{9}\alpha_{5})\sigma_{n}^{++}\sigma_{n}^{+-}+\frac{4}{9}\alpha_{5}(\sigma_{n}^{++}\sigma_{n}^{++}+\sigma_{n}^{+-}\sigma_{n}^{+-})) \nonumber\\
&+&\frac{4}{15}\alpha_{4}\sigma_{n}^{++}\sigma_{n}^{+-}+\frac{1}{15}\alpha_{4}(\sigma_{p}^{++}\sigma_{n}^{+-}+\sigma_{p}^{+-}\sigma_{n}^{++})\Big)F(x) \ .
\end{eqnarray}
Here $\alpha_{1}$=1.376, $\alpha_{2}$=1.002, $\alpha_{3}$=1.795, $\alpha_{4}$=0.300, $\alpha_{5}$=0.181.

\underline{Triple scattering}

As it was explained before in Appendix C the triple scattering term is very small. Its 0.8\% contribution to the total cross section of the unpolarized $|h \rangle$ off $^7$Li with $M_{J}$=1/2 gives a small correction to 11\% contribution to shadowing due to the double scattering term. Therefore, we will not give the analytical expression of the triple term but rather will take it into account in numerical results presented in Appendix E.

\section{Unpolarized scattering off polarized $^7$Li targets with $M_{J}$=3/2 or -3/2 and $M_{J}$=1/2 or -1/2}
\label{Appendix E}

In this appendix we will present the total scattering cross section of the unpolarized hadronic projectile $|h\rangle$ off a target of $^7$Li with $M_{J}=3/2$ and 1/2. We define the cross section off $^7$Li with $M_{J}$=3/2 as
\begin{equation}
\sigma_{A}^{3/2}=\frac{1}{2}\Big(\sigma_{A}^{+,3/2}+\sigma_{A}^{-,3/2}\Big)=\frac{1}{2}\Big(\sigma_{A}^{+,3/2}+\sigma_{A}^{+,-3/2}\Big) \ .
\end{equation}
Using our results of Appendices B and C and the relation
\begin{eqnarray}
\sigma_{p}^{++}+\sigma_{p}^{++}&=&\sigma_{eff} \ , \nonumber\\ 
\sigma_{n}^{++}+\sigma_{n}^{++}&=&\sigma_{eff} \ ,
\end{eqnarray}
we obtain for the ratio $\sigma_{A}^{3/2}/7\sigma_{eff}$, whose deviation from 1 describes nuclear shadowing,
\begin{eqnarray}
\frac{\sigma_{A}^{3/2}}{7\sigma_{eff}}&=&1-\frac{9}{28}\frac{\sigma_{eff}}{\pi(R^2+3B)}F(x)-\frac{\sigma_{eff}}{\pi(R^2+3B)}\Big(\frac{33}{15}\alpha_{1}+\frac{12}{15}\alpha_{3}\Big)\frac{1}{7}F(x) \nonumber\\
&-&\frac{\sigma_{eff}}{\pi R^2}\Big(\frac{28.3}{15}\alpha_{4}+\frac{32}{15}\alpha_{5}+\frac{12}{15}\alpha_{6}\Big)\frac{1}{7}F(x)+0.0061\,g(x) \ .
\end{eqnarray}
The last term is the triple scattering term. After having substituted all numerical factors we obtain
\begin{equation}
\frac{\sigma_{A}^{3/2}}{7\sigma_{eff}}=1-0.0922\,e^{-176 \cdot x^2}+0.0061\,g(x) \ .
\label{unpl32}
\end{equation}

The scattering cross section of the unpolarized $|h\rangle$ off the $^7$Li target with $M_{J}=-3/2$ is the same.

Similarly to the the above calculations and using Appendix D,  we can present the ratio $\sigma_{A}^{1/2}/7\sigma_{eff}$ for the target with $M_{J}$=1/2
\begin{eqnarray}
\frac{\sigma_{A}^{1/2}}{7\sigma_{eff}}&=&1-\frac{9}{28}\frac{\sigma_{eff}}{\pi(R^2+3B)}F(x)-\frac{\sigma_{eff}}{\pi(R^2+3B)}\Big(\frac{12}{15}\alpha_{1}+\frac{33}{15}\alpha_{3}\Big)\frac{1}{7}F(x) \nonumber\\
&-&\frac{\sigma_{eff}}{\pi R^2}\Big(\frac{9}{15}\alpha_{4}+\frac{72}{15}\alpha_{5}+\frac{12}{15}\alpha_{6}\Big)\frac{1}{7}F(x)+0.0083\,g(x) \ .
\end{eqnarray}
Numerically, this ratio is
\begin{equation}
\frac{\sigma_{A}^{1/2}}{7\sigma_{eff}}=1-0.1069\,e^{-176 \cdot x^2}+0.0083\,g(x) \ .
\label{unpl12}
\end{equation}

We used Eqs.\ (\ref{unpl32}) and (\ref{unpl12}) to find the scattering cross section of the unpolarized projectile off an unpolarized $^7$Li target.

\newpage
\begin{figure}
\centering
\mbox{\epsfig{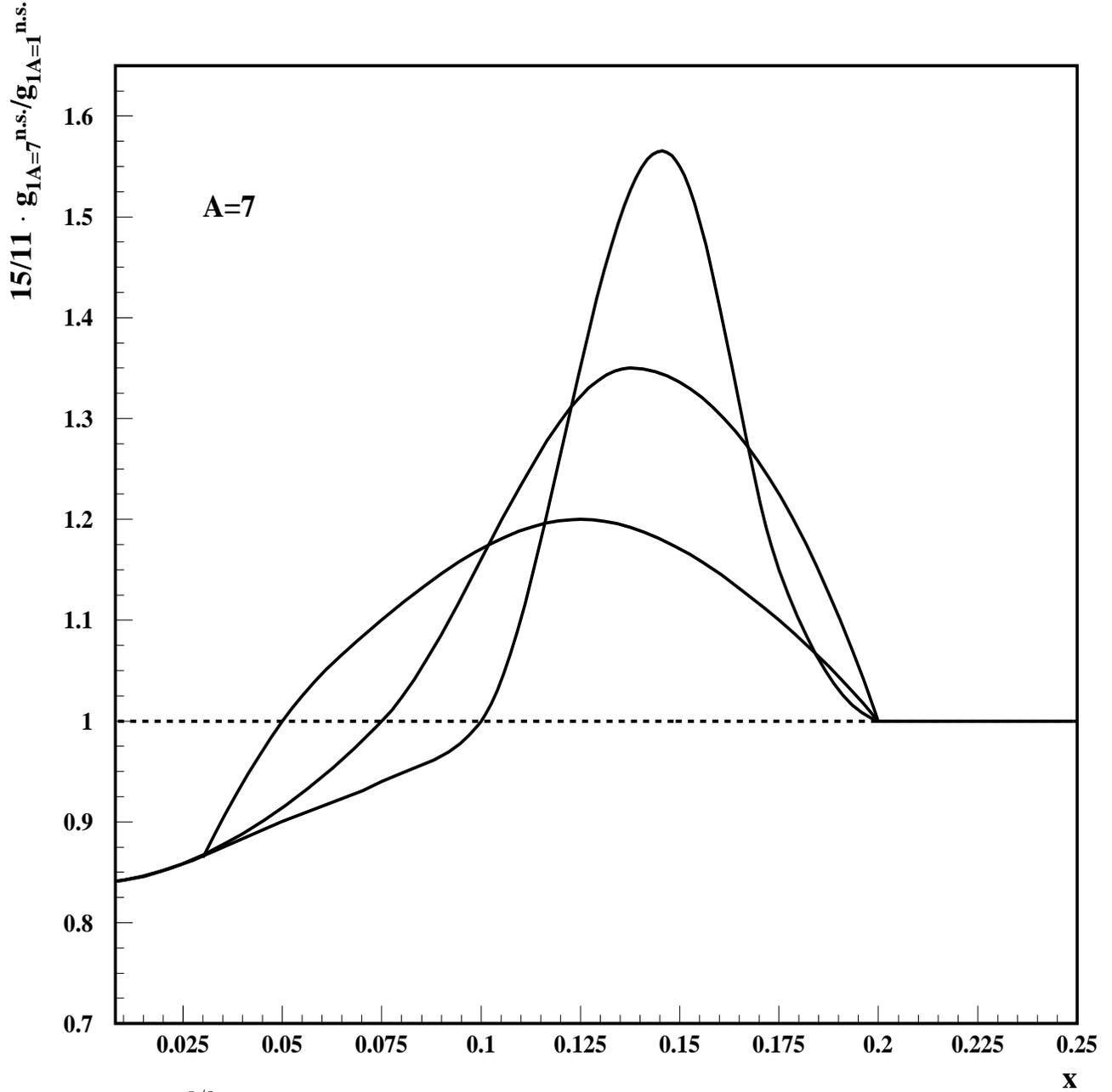}}
\caption{$g_{1A=7}^{n.s.\,3/2}(x,Q^2)/g_{1A=1}^{}(x,Q^2)$ as a function of $x$. The straight dashed line is the impulse approximation. The solid lines is a result of our calculation of shadowing and modeling of enhancement, which preserves R.}
\label{fig1}
\end{figure}

\newpage
\begin{figure}
\centering
\mbox{\epsfig{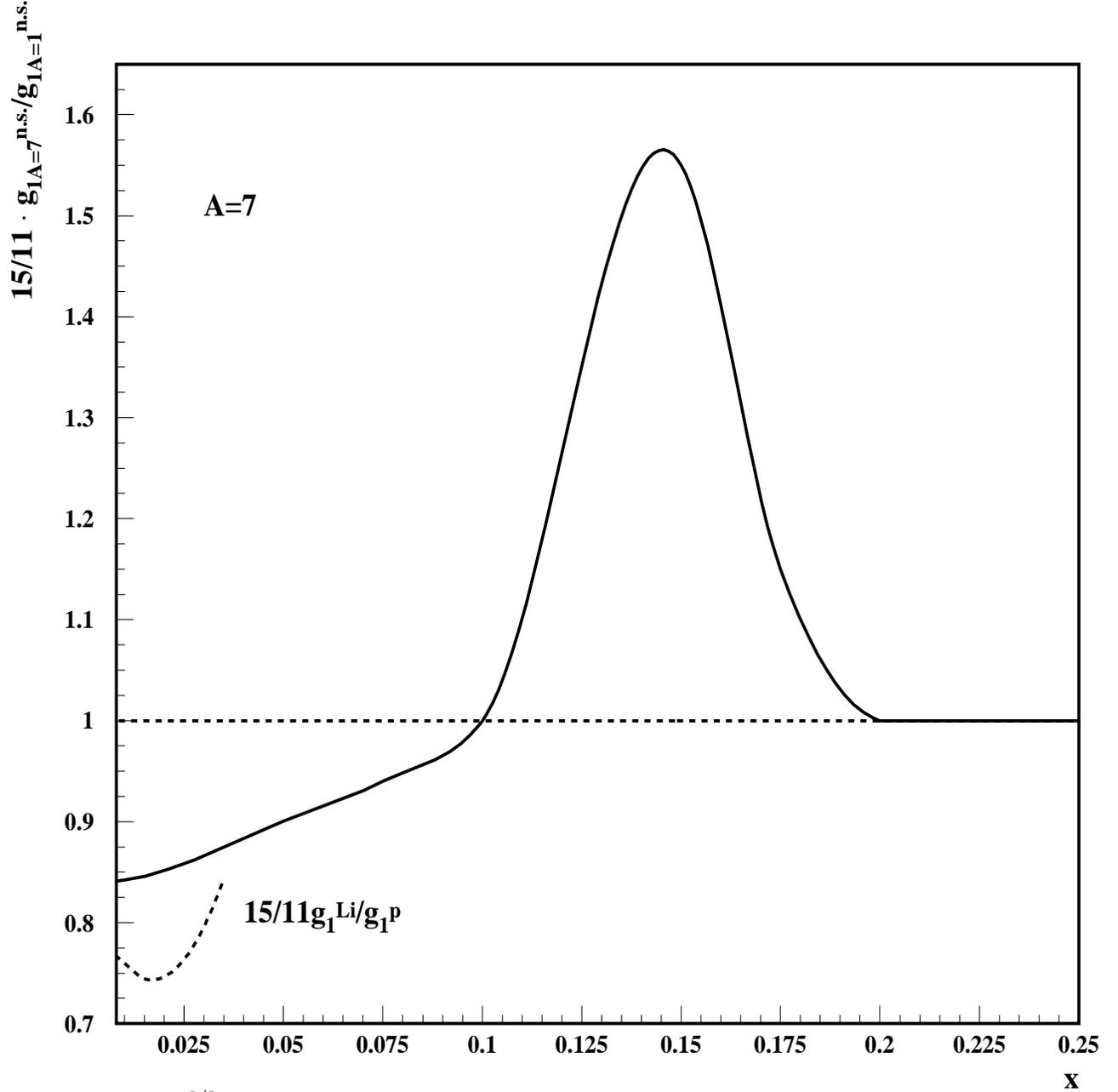}}
\caption{$g_{1A=7}^{n.s.\,3/2}(x,Q^2)/g_{1A=1}^{n.s.}(x,Q^2)$ as a function of $x$. The straight dashed line is the impulse approximation. The solid line is a result of our calculation of shadowing and modeling of enhancement, which preserves R. The curved dashed line is $g_{1}^{{\rm Li}\,3/2\,3/2}(x,Q^2)/g_{1}^{p}(x,Q^2)$.}
\label{fig2}
\end{figure}

\newpage
\begin{figure}
\centering
\mbox{\epsfig{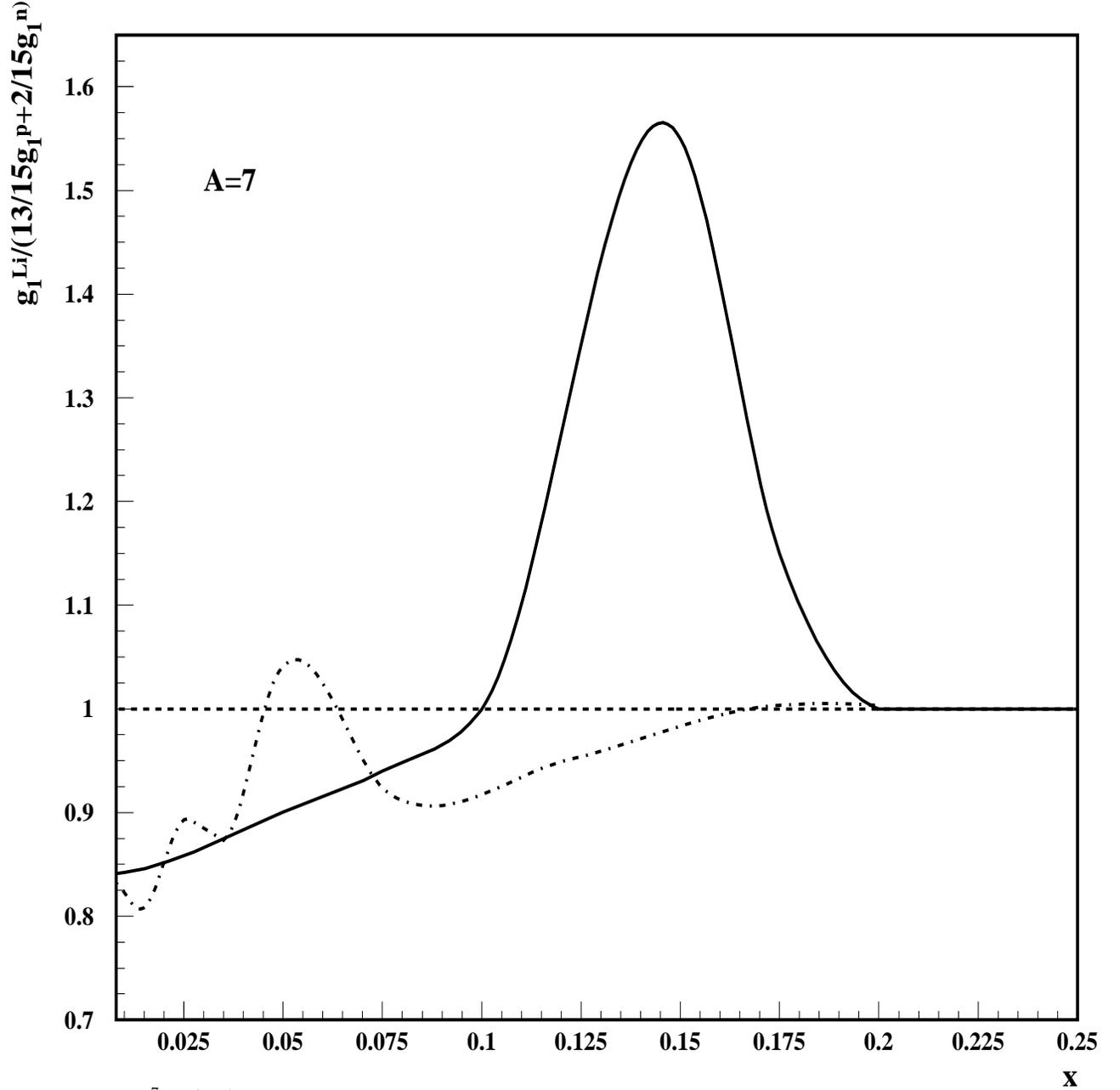}}
\caption{$g_{1}^{^{7}\rm{Li}\,3/2\,3/2}(x,Q^2)/(13/15\,g_{1}^{p}(x,Q^2)+2/15g_{1}^{n}(x,Q^2))$ as a function of $x$. The solid line is a result of our calculation of shadowing and modeling of enhancement, which preserves R. The dashed-dotted line is given by Eq.\ (\ref{ineq2}).}
\label{fig3}
\end{figure}

\newpage
\begin{figure}
\centering
\mbox{\epsfig{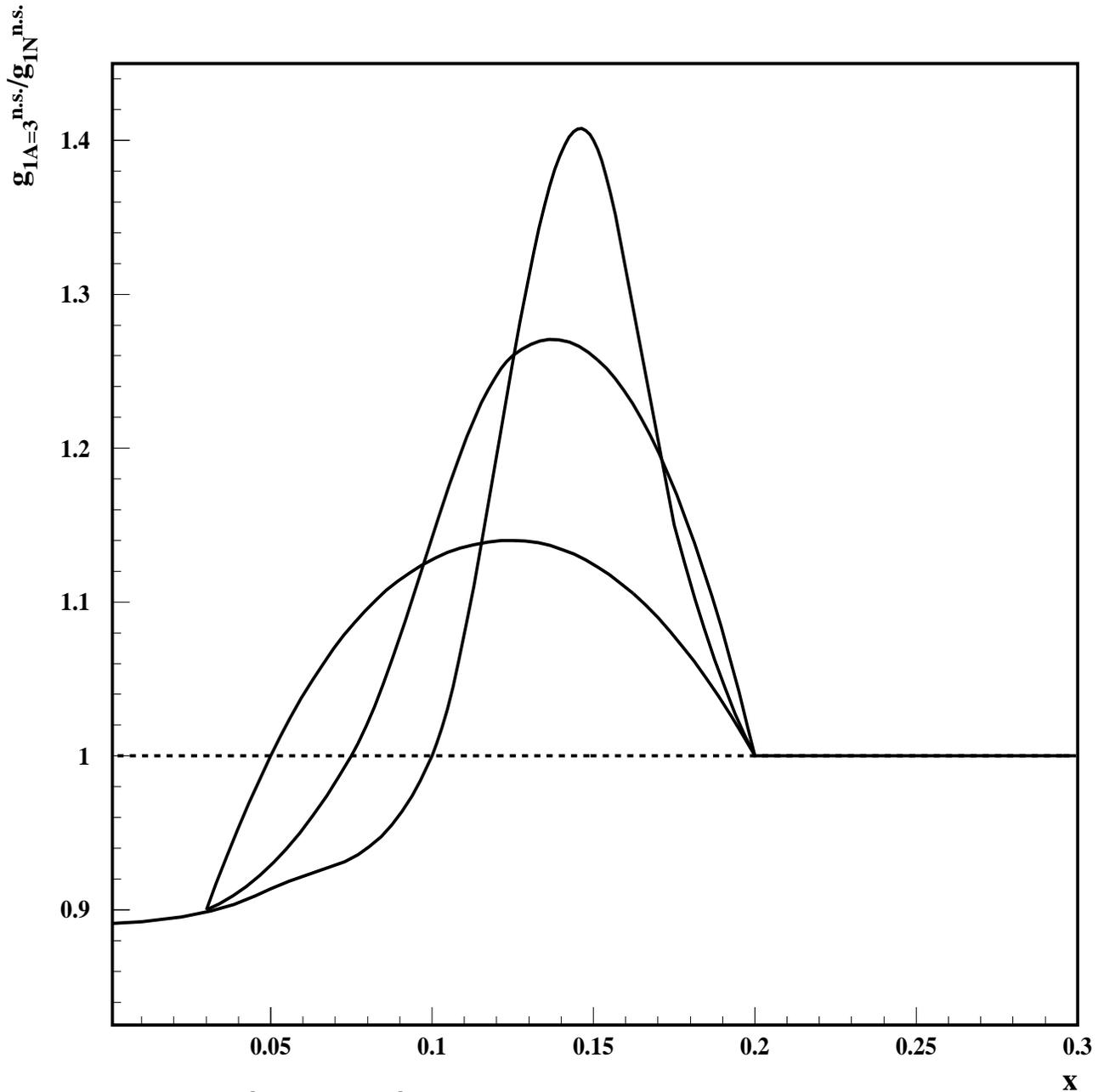}}
\caption{$g_{1A=3}^{n.s.}(x,Q^2)/g_{1A=1}^{}(x,Q^2)$ as a function of $x$. The straight dashed line is the impulse approximation. The solid lines is a result of our calculation of shadowing and modeling of enhancement, which preserves R.}
\label{fig4}
\end{figure}

\begin{figure}
\centering
\mbox{\epsfig{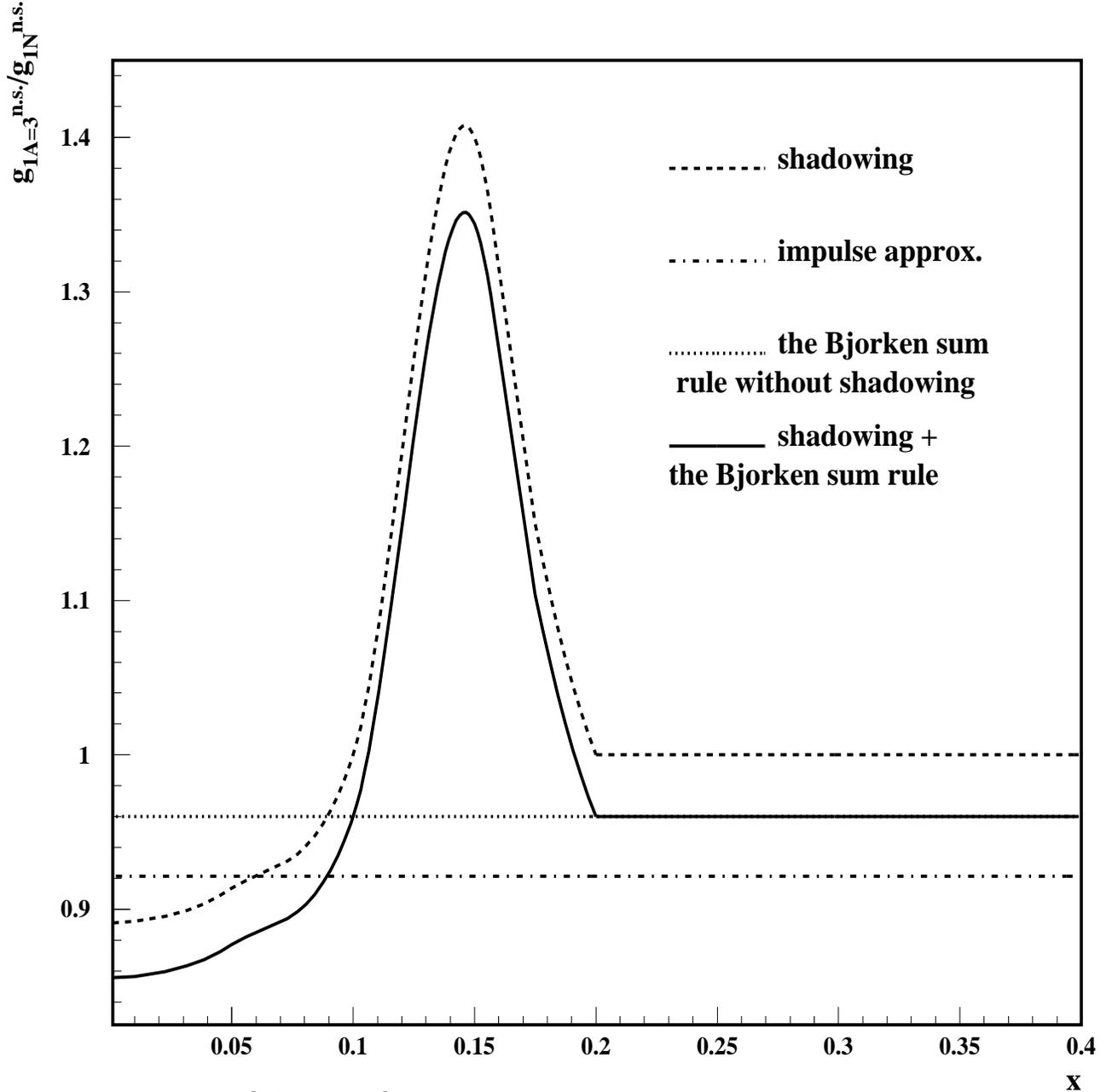}}
\caption{$g_{1A=3}^{n.s.}(x,Q^2)/g_{1N}^{n.s.}(x,Q^2)$ as a function of $x$. The dashed 
line represents nuclear shadowing at small $x$ and enhancement. The solid line is the result of the fit constrained to preserve the Bjorken sum rule.}
\label{fig5}
\end{figure}


\begin{thebibliography}{99}
\bibitem{gribov} V.N.~Gribov, Sov.J.Nucl.Phys. {\bf 9} 1969) 369; 
Sov.Phys.JETP {\bf 29} 1969, 483; ibid {\bf 30} (1970) 709.
\bibitem{FS981}L.~Alvero,  L. L.~Frankfurt and M.~Strikman,
hep-ph@xxx.lanl.gov - 9810331, Europ.J.Phys. in press;
 L. L.~Frankfurt and M.~Strikman,
hep-ph@xxx.lanl.gov - 9812322.
\bibitem{FS88} L.~Frankfurt and M.~Strikman, Phys. Rep. {\bf 160} 
 (1988) 235.
\bibitem{FSL} L.~Frankfurt, M.~Strikman, and S.~Liuti, Phys. Rev. Lett. {\bf 65} (1990) 1725.
\bibitem{Pirner}T.~Gousset and H.J.~Pirner,  Phys. Lett. {\bf B 375} (1996) 349. 
\bibitem{DY}D.M.~Alde, {\it et al.}, Phys. Rev. Lett. {\bf 64} (1990) 2479.
\bibitem{sf} C.~Ciofi degli Atti, E.~Pace, and G.~Salme, Phys. Rev. {\bf C 46} (1992) R1591; R.-W.~Schulze and P.U.~Sauer, Phys. Rev. {\bf C 48} (1993) 38; R.-W.~Schulze and P.U.~Sauer, Phys. Rev. {\bf C 56} (1997) 2293.
\bibitem{bsr} C.~Ciofi degli Atti, E.~Pace, and G.~Salme, Phys. Rev. {\bf C 48} (1993) R968; L.~Kaptari and A.~Umnikov, Phys. Lett. {\bf B 240} (1990) 203. 
\bibitem{FGS} L.~Frankfurt, V.~Guzey and M.~Strikman, Phys. Lett. {\bf B 381} (1996) 379. 
\bibitem{Piller} J.~Edelmann, G.~Piller, W.~Weise, 
Z. Phys. A357 (1997) 129, Phys. Rev. C57 (1998) 3392-3405. 
\bibitem{crabb}D.~Crabb, private communication.
\bibitem{Landau} L.D.~Landau and E.M.~Lifshitz ''Quantum mechanics, non relativistic theory'', Pergamon Press 1977.
\bibitem{ABFR} G.~Altarelli, R.~Ball, S.~Forte, and G.~Ridolfi, hep-ph@xxx.lanl.gov - 9803237. 
\bibitem{VDM}T.H.~Bauer, R.D.~Spital, D.R.~Yennie, F.M.~Pipkin,
 Rev.Mod.Phys. {\bf 50}
(1978) 261, ERRATUM-ibid. {\bf 51} (1979) 407.
\bibitem{Jaffe}  R.L.~Jaffe and A.~Manohar, Nucl. Phys. {\bf B 321} (1989) 343.
\bibitem{adeva} B.~Adeva, {\it et al.,} preprint CERN-EP/98-85.
\bibitem{quench} C.D.~Goodman and S.D.~Bloom in ''Spin Excitations in Nuclei'', p. 143,  ed. by F.~Petrovich {\it et al.,}, Plenum Press, 1984.
\bibitem{GW} J.F.~Germound and C.~Wilkin, Phys.Lett. {\bf B 59} (1975) 317.
\bibitem{Strikman} M.~Strikman, Proc.Symp. on Spin Structure of the Nucleon, Yale, 1994,
edited by V.H.~Hughes and C.~Cavata, World Scientific, 1995.
\bibitem{NS}N.N.~Nikolaev and W.~Sch\"afer, Phys.Lett. {\bf B 398} (1997) 254.



\end{thebibliography}
\end{document}